\documentclass[fleqn,usenatbib]{mnras}

\usepackage{newtxtext,newtxmath}
\usepackage{CJKutf8}

\usepackage[T1]{fontenc}
\defcitealias{prime1}{Primeval~I}
\defcitealias{prime2}{Primeval~II}
\defcitealias{prime3}{Primeval~III}
\defcitealias{prime4}{Primeval~IV}
\defcitealias{prime5}{Primeval~V}
\defcitealias{prime6}{Primeval~VI}
\defcitealias{prime7}{Primeval~VII}

\DeclareRobustCommand{\VAN}[3]{#2}
\let\VANthebibliography\thebibliography
\def\thebibliography{\DeclareRobustCommand{\VAN}[3]{##3}\VANthebibliography}

\usepackage{color, colortbl}
\definecolor{LRed}{rgb}{1,.9,.9}
\definecolor{LCyan}{rgb}{0.88,1,1}

\usepackage{graphicx}	
\usepackage{amsmath}	

\usepackage[usenames,dvipsnames]{xcolor}
\usepackage{soul}

\title[The first age benchmark L subdwarf]{Primeval very low-mass stars and brown dwarfs -- VIII. The first age benchmark L subdwarf, a wide companion to a halo white dwarf}
\author[Z. H. Zhang et al.]{Z. H. Zhang (\begin{CJK*}{UTF8}{gbsn}张曾华\end{CJK*}),$^{1,2}$\thanks{E-mail:
zz@nju.edu.cn} R. Raddi,$^{3}$ A. J. Burgasser,$^{4}$ S. L. Casewell,$^{5}$ R. L. Smart,$^{6}$ 
\newauthor 
M. C. G\'alvez-Ortiz,$^{7}$ H. R. A. Jones,$^{8}$ S. Baig,$^{8}$ N. Lodieu,$^{9,10}$ B. Gauza,$^{11}$ Ya. V. Pavlenko,$^{9,10,12}$ 
\newauthor 
Y. F. Jiao (\begin{CJK*}{UTF8}{gbsn}焦云帆\end{CJK*}),$^{1,2}$ 
Z. K. Zhao (\begin{CJK*}{UTF8}{gbsn}赵坤瑶\end{CJK*}),$^{1,2}$ 
S. Y. Zhou (\begin{CJK*}{UTF8}{gbsn}周思琰\end{CJK*})$^{1,2}$ and D. J. Pinfield$^{8}$ \\
$^{1}$School of Astronomy \& Space Science, Nanjing University, 163 Xianlin Avenue, Nanjing 210023, China \\
$^{2}$Key Laboratory of Modern Astronomy and Astrophysics, Nanjing University, Ministry of Education, Nanjing 210023, China \\
$^{3}$Universitat Polit\`ecnica de Catalunya, Departament de F\'isica, c/ Esteve Terrades 5, E-08860 Castelldefels, Spain \\
$^{4}$Department of Astronomy \& Astrophysics, University of California San Diego, La Jolla, CA 92093, USA \\
$^{5}$School of Physics and Astronomy
University of Leicester, University Road, Leicester, LE1 7RH, UK \\
$^{6}$Istituto Nazionale di Astrofisica, Osservatorio Astronomico di Torino, Strada Osservatrio 20, I-10025 Pino Torinese, Italy \\
$^{7}$Centro de Astrobiolog{\'i}a (CSIC-INTA), Ctra. Ajalvir km 4, E-28850 Torrejo{\'n} de Ardoz, Madrid, Spain \\
$^{8}$Centre for Astrophysics Research, University of Hertfordshire, Hatfield, Hertfordshire AL10 9AB, UK \\
$^{9}$Instituto de Astrof{\'i}sica de Canarias, E-38205 La Laguna, Tenerife, Spain \\
$^{10}$Universidad de La Laguna, Dept. Astrof{\'i}sica, E-38206 La Laguna, Tenerife, Spain \\
$^{11}$Janusz Gil Institute of Astronomy, University of Zielona G\'ora, Lubuska 2, PL-65-265 Zielona G\'ora, Poland \\
$^{12}$Main Astronomical Observatory, Academy of Sciences of the Ukraine, Golosiiv Woods, UA-03680 Kyiv-127, Ukraine
}

\date{Accepted 2024 July 26. Received 2024 July 23; in original form 2024 May 28}

\pubyear{\the\year{}}

\begin{document}
\label{firstpage}
\pagerange{\pageref{firstpage}--\pageref{lastpage}}
\maketitle

\begin{abstract}
We report the discovery of five white dwarf + ultracool dwarf systems identified as common proper motion wide binaries in the {\sl Gaia} Catalogue of Nearby Stars. The discoveries include a white dwarf + L subdwarf binary, VVV~1256$-$62AB,  a gravitationally bound system located 75.6$^{+1.9}_{-1.8}$~pc away with a projected separation of 1375$^{+35}_{-33}$ au.  
The primary is a cool DC white dwarf with a hydrogen dominated atmosphere, and has a total age of $10.5^{+3.3}_{-2.1}$ Gyr, based on white dwarf model fitting. 
The secondary is an L subdwarf with a metallicity of [M/H] = $-0.72^{+0.08}_{-0.10}$ (i.e. [Fe/H] = $-0.81\pm0.10$) and $T_{\rm eff}$ = 2298$^{+45}_{-43}$ K based on atmospheric model fitting of its optical to near infrared spectrum, and likely has a mass just above the stellar/substellar boundary. 
The sub-solar metallicity of the L subdwarf and the system's total space velocity of 406 km s$^{-1}$ indicates membership in the Galactic halo, and it has a flat eccentric Galactic orbit passing within 1~kpc of the centre of the Milky Way every $\sim$0.4Gyr and extending to 15--31 kpc at apogal.
VVV~1256$-$62B is the first L subdwarf to have a well-constrained age, making it an ideal benchmark of metal-poor ultracool dwarf atmospheres and evolution. 

\end{abstract}




\section{Introduction}
The L dwarf spectral classification \citep{kirk99,mart99} was established to classify brown dwarfs (BDs) and extremely low-mass stars (ELMS) with effective temperature $T_{\rm eff}$ $\approx$ 1200--2300 K \citep{kirk21}. 
L dwarfs are difficult to characterize individually because of the mass-age degeneracy at the boundary between stars and BDs, and are composed of
ELMS, transitional BDs \citep[][hereafter, \citetalias{prime3}]{prime3}, and degenerate BDs \citep[][hereafter, \citetalias{prime6}]{prime6}.
All of these source initially cool and dim over their first Gyr as they dissipate initial heat of formation \citep[e.g. fig 8 in][]{burr01}.
Stars with mass $\la 0.1 M_{\odot}$ ultimately reach a steady luminosity at late-M/early L spectral types, while transitional BDs (e.g. GD 165B; \citealt{kirk93,kirk99b}) are massive enough to have long-lasting, low-intensity hydrogen fusion that can slow down their cooling.
Degenerate BDs (e.g. Gl 229B, \citealt{naka95}; Kelu-1, \citealt{ruiz97}) comprise the majority of BDs and have no hydrogen fusion, and thus cool over the entire lifetimes.

L subdwarfs are the sub-solar metallicity counterparts of L dwarfs \citep{burg03,burg04}, and like M subdwarfs are classified into three metallicity classes: subdwarfs (sdL, [Fe/H] $\le -0.3$), extreme subdwarfs (esdL, $-1.7 <$ [Fe/H] $\le -1.0$), and ultra subdwarfs (usdL, [Fe/H] $\le -1.7$) \citep[][hereafter, \citetalias{prime1}]{prime1}. L subdwarfs exhibit distinct spectral features from L dwarfs, and have bluer optical to near infrared (NIR) colours (\citealt{gizi06,burg07,kirk10}; \citealt{prime4}, hereafter, \citetalias{prime4}). 
L subdwarfs are 200-400 K warmer than field L dwarfs with the same subtype \citepalias[e.g. fig. 4 in][]{prime3}.
The esdLs, usdLs, and a small fraction of sdLs are kinematically associated with the Galactic halo \citepalias[e.g. fig. 23 in][]{prime4}. Most sdLs are associated with the Galactic thick disc and L dwarfs with the thin disc. 

L subdwarfs are composed of metal-poor ELMS \citep[e.g. J0452$-$36B,][hereafter, \citetalias{prime7}]{prime7} and transitional BDs \citep[e.g. J0532+82;][]{burg03,burg08}. Metal-poor degenerate BDs are not within the L subdwarf sequence because after $\sim$ 10 Gyr they have already cooled to T or Y (sub)dwarf temperatures \citepalias{prime6}. 
L subdwarfs have a wider $T_{\rm eff}$ range but squeeze into a narrower mass range. Mid- to late-type L subdwarfs have a declining low rate of hydrogen fusion and are associated with a substellar transition zone \citepalias{prime3}. With a metallicity of [Fe/H] $\approx -2.4$, SDSS J0104+15 \citep[][hereafter, \citetalias{prime4}]{prime4} is the most metal-poor L subdwarf in the substellar transition zone known to date.

Field L dwarfs have diverse ages reflecting the Galactic disc population, spanning $\sim$0.5-8 Gyr \citep{redd06}. L subdwarfs likely have similar ages as other stars in the Galactic thick disc or halo \citep[$\sim$8-14 Gyr, e.g.][]{kili17}. 
However, we can not directly measure the age of an individual field L (sub)dwarf. Benchmark L (sub)dwarfs with known age are needed to test atmospheric and evolutionary models.
L (sub)dwarfs with wide white dwarf (WD) companions serve as ideal benchmarks in this capacity, as the age of the L companion is constrained by the WD's cooling age and progenitor's lifetime.
Eight WD + L/T/Y dwarf binaries are currently known \citep{beck88,stee09,dayj11,luhm11,deac14,zhan20,meis20,Fren23}, all of which belong to disc population given the absence of metal-poor features in the companion. 
No WD + L subdwarf wide binaries have yet been reported. 
J0452$-$36 AB (esdM1 + esdL0) is the only currently known star + L subdwarf wide binary \citepalias{prime7}, but the difficulty in inferring the precise ages of M (sub)dwarfs (e.g., \citealt{burg17}) implies this system is not a viable age benchmark.

The evolution of WDs is a well-understood cooling process in which the degenerate WD radiates its progenitor's core thermal energy directly to space. The atmospheric parameters of WDs ($T_{\rm eff}$ and $\log{g}$) can be measured with high precision with spectra, or both accurate photometry and parallax, and can be used to infer masses and cooling ages from WD cooling models \citep{althaus2010}. 
The mass of the WD remnant can then be related to the progenitor mass via the semi-empirical initial-to-final-mass relation \citep[e.g.][]{catalan2008}, which in turn provides the progenitor lifetime and the total age of the system. If the WD cooling age dominates the total age, as expected for old WDs, relatively accurate system ages can be inferred.

The Milky Way's halo population is expected to host the oldest WDs, which can reach relatively low temperatures ($T_{\rm eff}$ $\lesssim$ 3000~K) and thus exceedingly low luminosities ($L \lesssim$ 10$^{-5}$~L$_\odot$; \citealt{calc18}). The spectra of many WDs (e.g., DC spectral type), are featureless, making it difficult to measure their radial velocities (RVs), required to calculate UVW space velocities and assess halo membership. 
It is also not possible to directly infer the metallicity of a WD's progenitor. However, if a WD has a wide stellar/substellar companion, it is possible to determine the systems's RV, Galactic kinematics, and metallicity from the companion \citep[e.g.][]{raddi2022}. 

This is the eighth paper of a series titled {\sl Primeval very low-mass stars and brown dwarfs}, which presents discoveries, classification, characterization, and population properties of L and T subdwarfs.  
In this paper we present the discovery of five WD + ultracool dwarf (UCD) wide binaries, including the first halo WD + L subdwarf binary system VVV~1256$-$62AB, 
discovered in the {\sl Gaia} Catalogue of Nearby Stars \citep[GCNS;][]{smar21}. 
Section \ref{ssel} presents the selection of wide binaries from GCNS and criteria leading to the five discoveries reported here. 
Section \ref{swdmd} analyses four of the WD + UCD binaries, while section \ref{s1256ab} focuses on VVV~1256$-$62AB. 
Our conclusions are presented in Section \ref{scon}.

\section{Wide binary selection}
\label{ssel}
\begin{figure*}
	\includegraphics[width=\textwidth]{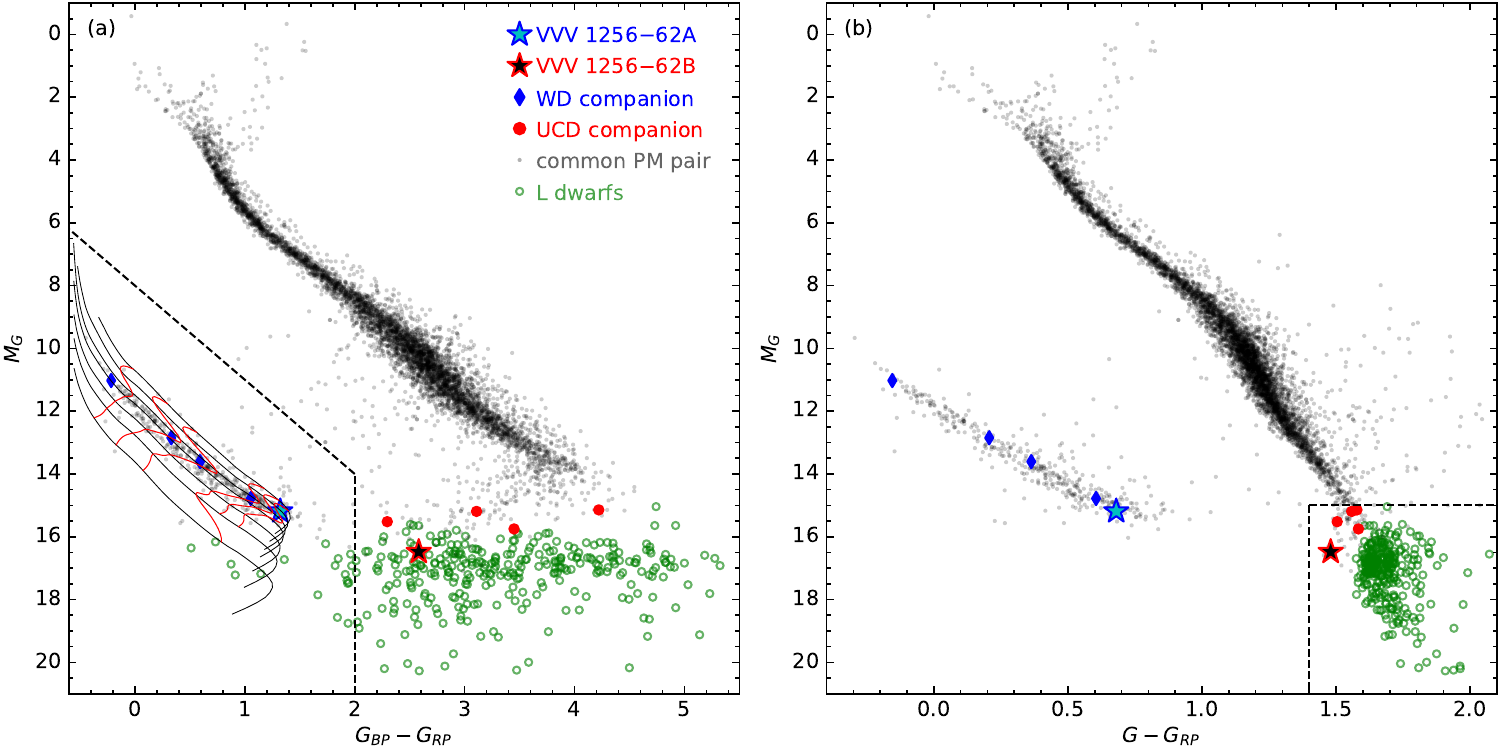}
    \caption{The H-R diagrams of 3,199 common PM pair with separations $\delta < 1^{\prime}$ in the GCNS. Four WD + late M dwarf binaries are highlighted with filled diamonds (WD) and filled circles (M dwarf). VVV~1256$-$62AB is highlighted with five-pointed stars. DA WD cooling tracks with masses of 0.3, 0.4, 0.5, 0.6, 0.8, 1.0, 1.2 $M_{\odot}$ (from upper right to lower left) and isochrones at 0.3, 1, 2, 5, 7, 9 Gyr (from upper left to lower right) are over plotted in the left diagram \citep{althaus2013,camisassa2016,camisassa2019} assuming a pure hydrogen atmosphere \citep{tremblay2013}. 
    }
    \label{fhrd}
\end{figure*}

\begin{figure}
	\includegraphics[width=\columnwidth]{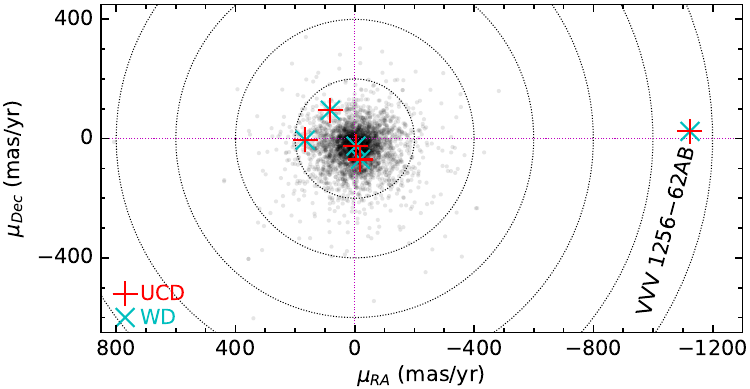}
    \caption{The PMs of 3,199 common PM pairs with separations $\delta < 1^{\prime}$ in the GCNS (grey dots). The difference in PM between components is less than 1 mas~yr$^{-1}$, ten times smaller than the symbol size.
    The five WD  + UCD pairs are indicated by crosses, with VVV~1256$-$62AB specifically labelled.
    }
    \label{fpm}
\end{figure}

\begin{figure*}
	\includegraphics[width=\textwidth]{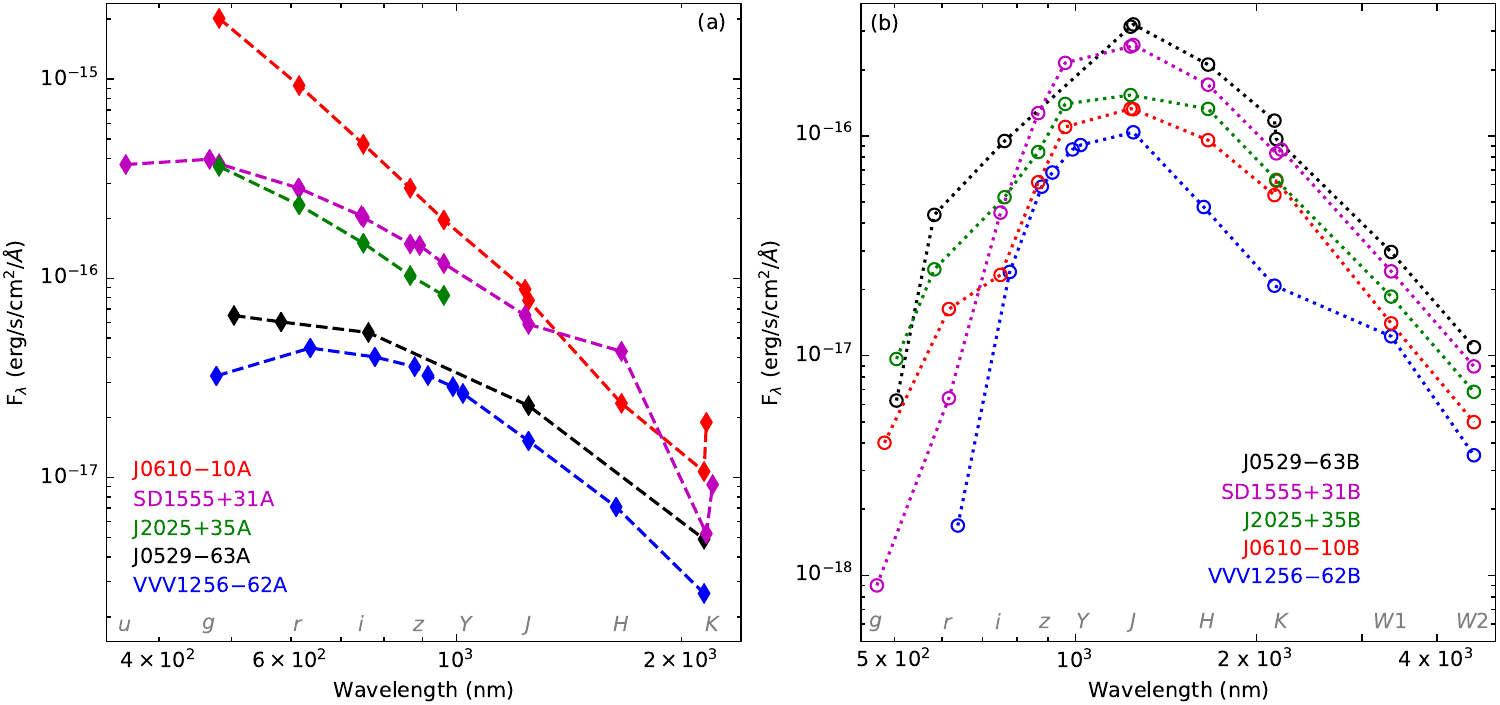}
    \caption{Optical to infrared SEDs of the five WD + UCD wide binaries in Table~\ref{tpm} based on {\sl Gaia}, SDSS \citep{york00}, DECaPS \citep{decaps1}, Two Micron All Sky Survey \citep[2MASS;][]{skru06}, VISTA VVV \citep{minn10}, UKIRT Hemisphere Survey \citep[UHS;][]{dye18}, and WISE \citep{wrig10} photometry. 
    WD SEDs are show in the left panel, UCD SEDs are shown in the right panel. Note {\sl Gaia} photometry is not quite accurate for faint UCDs (e.g. scattered $G_{\rm BP} - G_{\rm RP}$ colours of known L dwarfs in Fig. \ref{fhrd}). 
    SD1555+31B and VVV~1256$-$62B have spectroscopic confirmations and are discussed further in Sections \ref{ssd1555} and \ref{s1256ab}, respectively.}
    \label{fsed}
\end{figure*}

Precise proper motions (PM) of nearby stars measured by {\it Gaia} \citep{Gaia16,gaia23} provide a powerful tool to identify wide binaries in the solar neighbourhood (e.g., \citealt{elba21}). 
We searched for UCD (spectral types $\ge$ M7; masses $\lesssim$ 0.1 M$_{\odot}$) wide companions to WDs within the GCNS (331,312 sources). 
We constrained our search to systems with separations $\delta$ < 1 arcmin, with robust PMs ($\mu > 2$ mas/yr and $\mu/\mu_{error} > 4.5$) and tight constraints on common PM:
\begin{eqnarray}
    \label{eq:pmra}
    |\mu_{\rm RA1} - \mu_{\rm RA2}| < 1~{\rm mas~yr^{-1}} \\
    \label{eq:pmdec}
    |\mu_{\rm Dec1} - \mu_{\rm Dec2}| < 1~{\rm mas~yr^{-1}} 
\end{eqnarray}
We identified 3,199 common PM pairs with separation between 3 and 60 arcsec. 
Figs~\ref{fhrd} and~\ref{fpm} show the H-R diagrams and PMs of these
systems.
To narrow our sample to potential WD + UCD binaries, 
we used the $G_{BP}-G_{RP}$ versus $M_G$ H-R diagram (Fig.~\ref{fhrd}a) to select WD components, using criteria:
\begin{eqnarray}
    M_{\rm G} > 3(G_{\rm BP}-G_{\rm RP})+8 \\
    G_{\rm BP}-G_{\rm RP} < 2.0
\end{eqnarray}
UCDs are better detected in the {\sl Gaia} $G$ and $G_{RP}$ bands compared to in $G_{BP}$ band. Therefore, we used the $G-G_{RP}$ versus $M_G$ H-R diagram (Fig.~\ref{fhrd}b) to select UCD components using criteria:
\begin{eqnarray}
    M_{\rm G} > 15 \\
    G-G_{\rm RP} > 1.4
\end{eqnarray}
Only five of the 3,199 common PM pairs satisfied all of these criteria.
Table \ref{tpm} lists the {\sl Gaia} astrometry, separation, and spectral types of these pairs, while 
Figure~\ref{fsed} shows their optical to infrared spectral energy distributions (SED).

\begin{table*}
 \centering
  \caption[]{{\sl Gaia} Astrometry of five WD + UCD common PM pairs.}
\label{tpm}
  \begin{tabular}{l c r r rr rr}
\hline
Gaia DR3 & Other name & Separation  & Distance~~~ & $\mu_{\rm RA}$~~~ & $\mu_{\rm Dec}$~~~ & SpT & Ref. \\
&& (arcsec)  & (pc)~~~~~ & (mas yr$^{-1}$) & (mas yr$^{-1}$) && \\
\hline
4757030391786232576 & GALEX J052933.2$-$635653 & & 71.2 $\pm1.0$  & $166.1\pm0.3$ & $-4.9\pm0.2$ & WD & (1) \\ 
\smallskip
4757030327366948608 & 2MASS J05294026$-$6357091 & 47.92 & 70.8 $\pm1.3$  & $166.0\pm0.4$ & $-4.0\pm0.3$ & M8$^a$ & (2) \\
\hline
3004346292322852096 &2MASS J06100403$-$1031036 & & 100.8 $\pm0.6$ & $-4.0\pm0.1$ & $-25.1\pm0.1$  & WD & (3) \\ 
\smallskip
3004346257963119872 & 2MASS J06100078$-$1031393 & 59.95 & 93.3$^{+11.9}_{-9.5}$  & $-3.7\pm1.2$ & $-25.1\pm1.1$ & M8.5$^b$ &  (4) \\
\hline
5863122429179888000 &VVV J125644.42$-$620208.1  & & 75.6$^{+1.9}_{-1.8}$ & $-1124.0\pm0.3$ & $24.1\pm0.3$  & DC WD & (4,5) \\
\smallskip
5863122429178232704 & VVV J125641.09$-$620203.8 &  18.20 & 70.4$^{+7.3}_{-6.1}$ & $-1123.8\pm1.0$ &  $24.6\pm1.7$ & sdL3 & (6) \\
\hline
1321738565727229056 & GALEX J155516.9+315307 & & 59.8 $\pm0.3$ & $82.0\pm0.1$ &  $95.3\pm0.1$ & DA WD & (7) \\
\smallskip
1321738561431758592 & SDSS J155517.36+315316.8 & 10.86 & 59.1$^{+1.1}_{-1.0}$  & $82.2\pm0.3$ & $94.6\pm0.3$ & sdM9.5$^a$ & (4) \\
\hline
2056440344314369280 & IPHAS J202533.21+351509.6 & & 91.1 $\pm0.6$ & $-18.9\pm0.1$ & $-69.4\pm0.1$ & WD & (3) \\
\smallskip
2056440275586819456 & 2MASS J20253351+3515035 &  7.40 & 92.2$^{+3.8}_{-3.5}$ & $-18.6\pm0.4$ & $-70.0\pm0.5$ & M8.5$^b$ & (4) \\
\hline
\end{tabular}
\begin{list}{}{}
\item[References.] 
(1) \citet{gent19}; 
(2) \citet{reyl18};
(3) \citet{jime18};
(4) This paper; 
(5) \citet{gent21}
(6) \citetalias{prime5};
(7) \citet{kili20}. 
\item[Notes.] $^a$Photometric spectral type based on {\sl Gaia} photometry. $^b$Photometric spectral types based on Pan-STARRS photometry. 
\end{list}
\end{table*}

\begin{table*}
 \centering
  \caption[]{Properties of VVV~J12564163$-$6202039AB wide binary systems. }
\label{tprop}
  \begin{tabular}{l c c l}
\hline
Parameter & VVV~1256$-$62A &  VVV~1256$-$62B & Ref.   \\	
\hline 
Gaia DR3 & 5863122429179888000 & 5863122429178232704 & (1) \\
Spectral type &  DC WD & sdL3 & (2,3) \\
 $\alpha$ (2016)   & $12^{\rm h}56^{\rm m}43\fs49$ & $12^{\rm h}56^{\rm m}40\fs97$  & (1)  \\
 $\delta$ (2016)   &  $-62\degr02\arcmin08\farcs1$ & $-62^{\rm h}02^{\rm m}03\fs9$   & (1)   \\
$G_{\rm BP}$ &  20.22 $\pm$ 0.04 &  21.8 $\pm$ 0.2 & (1) \\
$G$ &  19.581 $\pm$ 0.003 & 20.701 $\pm$ 0.009 & (1) \\
$G_{\rm RP}$ &  18.90 $\pm$ 0.03 &  19.24 $\pm$ 0.05 & (1) \\
RUWE & 1.034 & 1.259 & (1) \\
$g$ (DECaPS) &   20.407 $\pm$ 0.008 & ---  & (4) \\
$r$ (DECaPS) &   19.429 $\pm$ 0.003 &  22.99 $\pm$ 0.39  & (4)\\
$i$ (DECaPS) &   19.12 $\pm$ 0.02 & 19.68 $\pm$ 0.03   & (4)\\
$z$ (DECaPS) &   19.00 $\pm$ 0.02 &  18.20 $\pm$ 0.01  & (4)\\
$Y$ (DECaPS) &   18.96 $\pm$ 0.02  & 17.77 $\pm$ 0.03  & (4) \\
$Z_V$ (2010.24) &  18.47 $\pm$ 0.03 & ---  & (5) \\
$Y_V$ (2010.24) &  18.27 $\pm$ 0.05 & ---  & (5) \\
$J_V$ (2010.18) &  18.13 $\pm$ 0.06 & ---  & (5) \\
$H_V$ (2010.18) &  17.96 $\pm$ 0.11 & ---  & (5) \\
$Ks_V$ (2010.18) &  18.03 $\pm$ 0.26 & ---  & (5)\\
$Z_V$ (2015.32) & 18.46 $\pm$ 0.05 &  17.93 $\pm$ 0.03  & (5) \\ 
$Y_V$ (2015.32) &  18.38 $\pm$ 0.06 & 17.04 $\pm$ 0.02  & (5) \\
$J_V$ (2015.40) &  18.21 $\pm$ 0.07 & 16.13 $\pm$ 0.01  & (5) \\
$H_V$ (2015.40) &  18.00 $\pm$ 0.14 & 15.94 $\pm$ 0.02  & (5) \\
$Ks_V$ (2015.40) &  --- & 15.78 $\pm$ 0.03  & (5) \\
$W1$ ({\sl CatWISE}) &  --- & 14.56 $\pm$ 0.02  & (6) \\
$W2$ ({\sl CatWISE}) &  --- & 14.59 $\pm$ 0.03  & (6) \\
$\varpi$ (mas) & 13.24 $\pm$ 0.33 & 14.20 $\pm$ 1.36  & (1) \\
Distance (pc)  & 75.6$^{+1.9}_{-1.8}$ &  70.4$^{+7.3}_{-6.1}$   & (1) \\
$\mu_{\rm RA}$ (mas yr$^{-1}$) &   $-$1124.0 $\pm$ 0.3 & $-$1123.8 $\pm$ 1.0    & (1) \\
$\mu_{\rm Dec}$ (mas yr$^{-1}$) &  24.1 $\pm$ 0.3 & 24.6 $\pm$ 1.7  & (1) \\
RV (km s$^{-1}$) & --- & $-$46.4 $\pm$ 1.9  & (3)  \\ 
$T_{\rm eff}$ (K) & $4440 \pm 250$  & 2298$^{+45}_{-43}$  & (3,7)  \\
log $g$ & 7.86 $\pm$ 0.05 & 5.43$^{+0.26}_{-0.15}$ & (3,7) \\
R ($R_{\sun}$) & --- & 0.089$^{+0.004}_{-0.003}$ & (3,7) \\
${\rm [M/H]}$ & --- &  $-$0.72$^{+0.08}_{-0.10}$ & (3,7) \\
${\rm [\alpha/Fe]}$ & --- & +0.13$^{+0.02}_{-0.02}$ & (3,7)  \\
${\rm [Fe/H]}$ & --- & $-0.81\pm0.10$ & (3,7) \\
Mass (M$_{\sun}$) & 0.62 $\pm$ 0.04 & 0.082 $\pm$ 0.001 & (3) \\
$\tau_{\rm cool}$ (Gyr) & $8.5 \pm 1.9$ & ---  & (3) \\
$\tau_{\rm prog.}$ (Gyr) & $2.0^{+1.8}_{-0.8}$ & ---  & (3) \\
$\tau_{\rm total}$ (Gyr) & $10.5^{+3.3}_{-2.1}$ & ---  & (3) \\
\hline 
Separation (arcsec) &  \multicolumn{2}{c}{18.2}   & (1)    \\
Proj. sep. (au) &  \multicolumn{2}{c}{1375$^{+35}_{-33}$}   & (1)   \\
Period (yr) &  \multicolumn{2}{c}{$\goa 6\times10^{4}$}   & (3)   \\
Proj. sep. ($r_{\rm J}$)$^a$ &  \multicolumn{2}{c}{$5.56\times10^{-3}$}   & (3)  \\
$-U$ (J) &  \multicolumn{2}{c}{$6.5\times10^{34}$} & (3)   \\
$V_{\rm tan}$ (km s$^{-1}$) & \multicolumn{2}{c}{$403\pm10$}  & (1)  \\
$U^b$ (km s$^{-1}$) & \multicolumn{2}{c}{$-361.3\pm8.3$}  & (3) \\ 
$V$ (km s$^{-1}$) & \multicolumn{2}{c}{$-183.6\pm5.6$}  & (3) \\ 
$W$ (km s$^{-1}$) & \multicolumn{2}{c}{$16.2\pm0.3$}   & (3) \\ 
$V_{\rm total}$ (km s$^{-1}$) & \multicolumn{2}{c}{$406\pm10$}   & (3) \\ 
\hline
\end{tabular}
\begin{list}{}{}
\item[]$^a$$r_{\rm J}$ is the Jacobi radius, the boundary where the Galactic tidal field exceeds the gravitational attraction of a wide binary \citep{jian10}. 
\item[]$^b$Positive toward the Galactic centre.
\item[]References:
(1) Gaia DR3 \citep{gaia23}
(2) \citet{gent21}; 
(3) This paper;
(4) DECaPS DR2 \citep{decaps1,decaps2}; 
(5) Vista VVV \citep{minn10};
(6) CatWISE2020 \citep{maro21};
(7) SAND atmosphere models \citep{Alvarado_2024};
\end{list}
\end{table*}

\begin{figure*}
	\includegraphics[width=\textwidth]{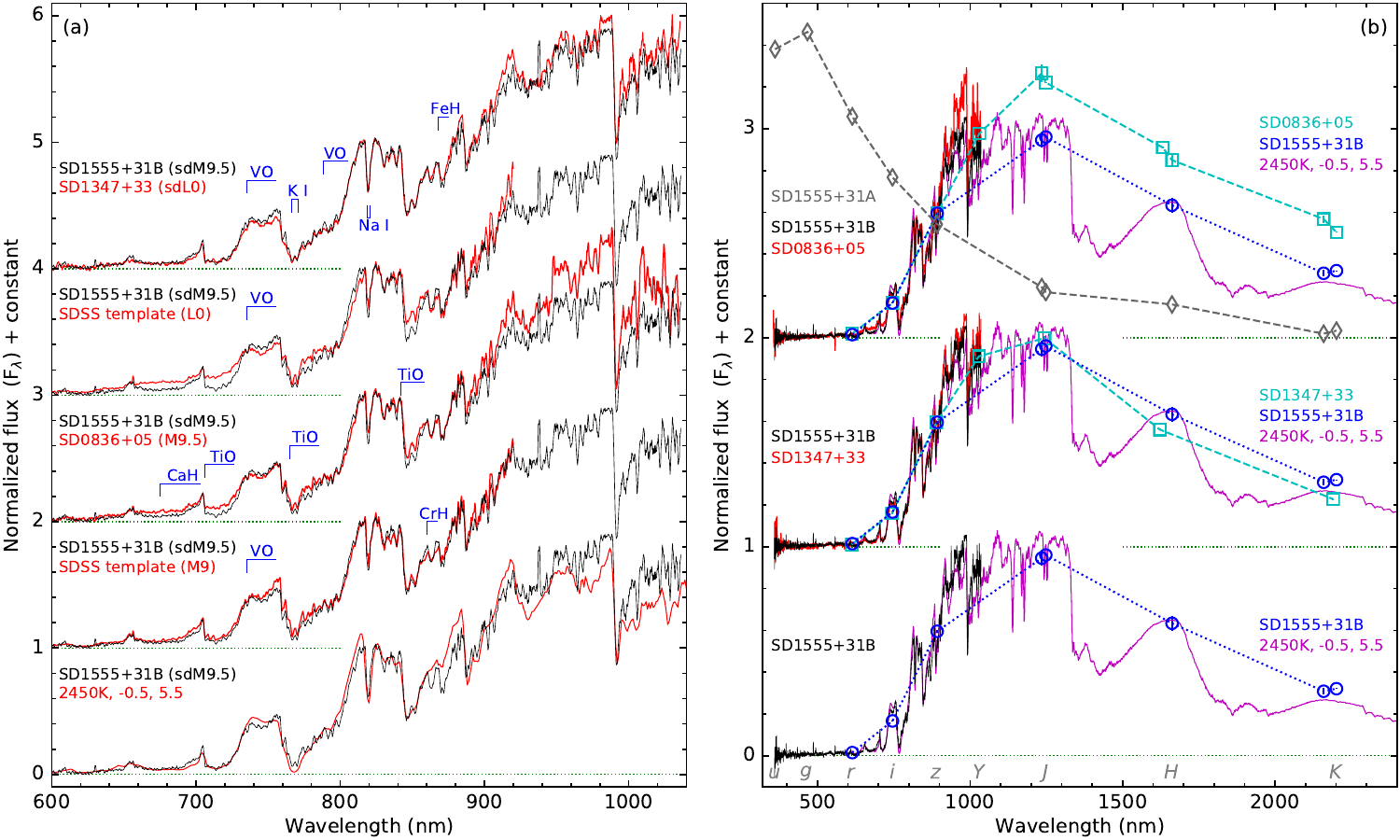}
    \caption{Left: The SDSS spectrum of SD1555+31B compared to SDSS M9 and L0 templates \citep{boch07}, the SDSS spectra of SD1347+33 (sdL0, \citetalias{prime1}) and SD0836+05 (M9.5; \citealt{west08}), and a BT-Settl model spectrum \citep{alla14} with $T_{\rm eff}$ = 2450K, [Fe/H] = $-$0.5, log $g$ = 5.5. The spectra of SD1555+31B and SD0836+05 are smoothed by 3 pixels. Right: The optical to NIR SED of SD1555+31A\&B (re-scaled by the same factor) compared to those of SD1347+33 and SD0836+05, and a BT-Settl model spectrum as in the left panel.
    }
    \label{fsd1555}
\end{figure*}

\section{Wide WD + M dwarf/subdwarf binaries}
\label{swdmd}
\subsection{A WD + sdM9.5 wide binary}
\label{ssd1555}
The white dwarf primary Gaia DR3 1321738565727229056 (WD4 in Table \ref{tpm}), aka SDSS~J155516.95+315307.2 (SD1555+31A) is classified as a DA WD by \citet{kili20}. It has a mass M = $0.553\pm0.008M_{\odot}$, $T_{\rm eff}$ = 6546 $\pm$ 28 K, and log $g$ = 7.936 $\pm$ 0.010 according to \citet{gent19}. SD1555+31A is at a distance of 59.8$\pm$0.3~pc. The UCD companion, Gaia DR3 1321738561431758592 (UCD4), i.e. SDSS J155517.36+315316.8 (SD1555+31B) is separated by 10$\farcs$86 from SD1555+31A at an equivalent distance of 59.1$^{+1.1}_{-1.0}$~pc, corresponding to a projected separation of 649$\pm$3 au. 

The optical spectrum of SD1555+31B was observed by the Sloan Digital Sky Survey \citep[SDSS;][]{york00} on 2011 July 6. Figure~\ref{fsd1555} shows this spectrum compared to SDSS spectra of the M9.5 dwarf SDSS~J083646.34+052642.6 (SD0836+05; \citealt{west08}), the sdL0 subdwarf SDSS J134749.74$+$333601.7 \citepalias[SD1347+33;][]{prime1}, and SDSS M9 and L0 dwarf templates from \citet{boch07}. 
SD1555+31B has similar spectrum to the SDSS M9 and L0 templates. 
The strength of its 745nm VO absorption, an indicator of spectral type M/L (sub)dwarfs boundary\citep{kirk14} is intermediate between these templates.
Indeed, SD1555+31B matches well with the M9.5 dwarf SD0836+05 at 745~nm VO band and at the 800-840~nm peak, which indicates a spectral type of M9.5. 
However, SD1555+31B has a relatively deep 863nm CrH absorption and stronger CaH (690nm) and TiO (715, 775, 850nm) bands, all signatures of mildly metal-poor UCDs (\citealt{burg07}, \citetalias{prime1}). 
The spectral profile of SD1555+31B matches well with the sdL0 SD1347+33, but has more flux in the 734-759~nm region, suggesting a slightly earlier spectral type. 
The optical and NIR photometric SED of SD1555+31B (Fig.~\ref{fsd1555}) is also similar to that of the sdL0 SD1347+33, with significant suppression in the NIR compared to the M9.5 SD0836+05 likely driven by enhanced H$_2$ collision induced absorption \citep{lins69,burg03}. This is again consistent with a sub-solar metallicity; hence, we classify SD1555+31B as an sdM9.5 subdwarf.

\subsection{Three additional WD + late M dwarf wide binaries}
The other three UCD companions (Table \ref{tpm}) are all late M dwarfs.  
The UCD candidate, Gaia DR3 4757030327366948608 (J0529$-$63B) was photometrically classified as an M8 dwarf based on optical and NIR photometry \citep{reyl18}. Its primary Gaia DR3 4757030391786232576 (J0529$-$63A), is listed as a WD in \citet{gent19} with a mass $M$ = 0.52-0.56 M$_{\odot}$ and $T_{\rm eff}$ = 4983-5042 K. J0529$-$63A is the second faintest of the five WD primaries in Fig. \ref{fhrd} with $M_G = 14.78$. Its optical and NIR SED (Fig. \ref{fsed}) is consistent with a cool WD. 

The two UCD candidates Gaia DR3 3004346257963119872 (J0610$-$10B) and Gaia DR3 2056440275586819456 (J2025+35B) both have photometric spectral types of M8.5 estimated from their $i-z$, $z-y$, $i-y$ colours of the Panoramic Survey Telescope and Rapid Response System \citep[Pan-STARRS;][]{cham16} and colour - spectral type correlations \citep{best18}. The WD primary of the first source, Gaia DR3 3004346292322852096 (J0610$-$10A) has a mass $M$ = 0.65-0.7 M$_{\odot}$ and $T_{\rm eff}$ = 19,571-20,247 K  \citep{gent19}. The WD primary of the second source, Gaia DR3 2056440344314369280 (J2025+35A) has a mass $M$ = 0.55-0.61 M$_{\odot}$ and $T_{\rm eff}$ = 8315-8509 K \citep{gent19}.

\section{A halo WD + L3 subdwarf wide binary}
\label{s1256ab}
The WD Gaia DR3 5863122429179888000 and its UCD companion Gaia DR3 5863122429178232704 (VVV J125641.09$-$620203.8) have the highest PMs among all 3,199 common PM pairs identified in this study (Fig. \ref{fpm}). Figure~\ref{ffc} shows the NIR field of the pair, which are separated by 18.2 arcsec, from Visible and Infrared Survey Telescope for Astronomy (VISTA) Variables in the Via Lactea \citep[VVV;][]{minn10} over the time-scale 2010 to 2015. Their PM over a baseline of $\sim$5 years is significant and visible by eye.
VVV~1256$-$62B was originally identified from the VVV Infrared Astrometric Catalogue \citep[VIRAC;][]{smit18} as a high PM source (1.1 arcsec~yr$^{-1}$) and classified as an sdL7 by \citet{smit18} based on a low signal-to-noise ratio infrared spectrum. \citet[][hereafter, \citetalias{prime5}]{prime5} subsequently reclassified this source as a sdL3 subdwarf based on higher quality data from the X-shooter spectrograph \citep{vern11} on the Very Large Telescope (VLT), shown in Figure~\ref{fspec}.

\subsection{Photometric observations}
\begin{figure}
\includegraphics[width=\columnwidth]{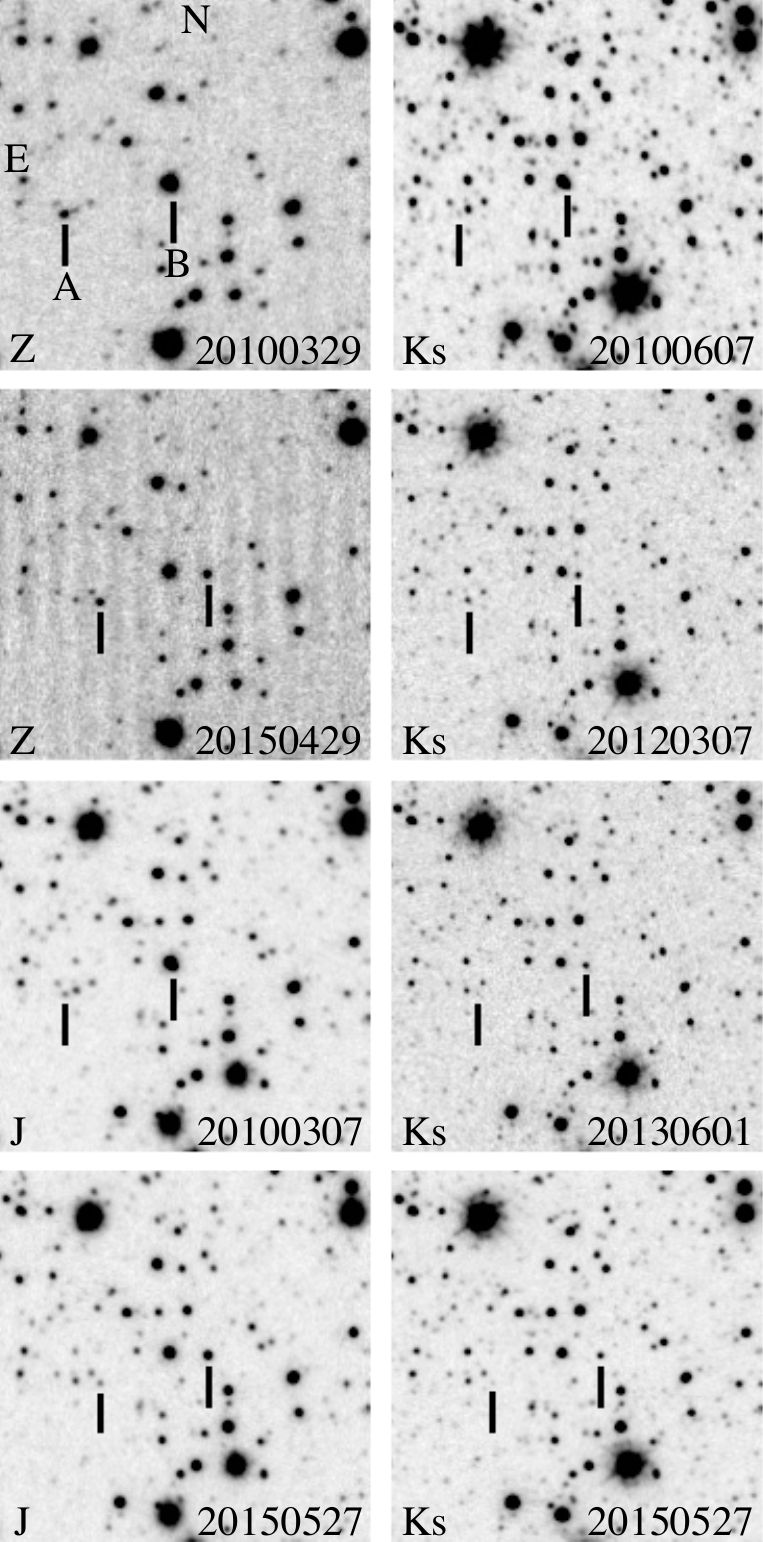}
    \caption{The VVV $Z$, $J$, and $K_s$ images of the field around VVV~1256$-$62A and B, indicated by the vertical bars (6 arcsec  in length) below each source. All images have a size of 1 arcmin with north up and east left. 
    The filter name and observation dates in yyyymmdd format are labelled at the bottom of each image.
    Note that the primary is significantly fainter than the secondary in all bands.}
    \label{ffc}
\end{figure}

VVV~1256$-$62A and B have optical and NIR photometric observations from 
the {\sl Gaia} mission, 
the DECam Plane Survey \citep[DECaPS;][]{decaps1,decaps2}, 
the VVV (Table \ref{tprop}),
and CatWISE2020 \citep{maro21};
The VVV images in Fig.~\ref{ffc} show that VVV~1256$-$62B was blended with a brighter background object in the 2010 images.  
The WD companion VVV~1256$-$62A has poor $Ks$ band detections in the VVV;
it was next to a brighter background object and likely not detected in the 2012 Ks band image, and was 
barely detected in 2013 and 2015 $Ks$-band images.
It was therefore not picked up VIRAC which is based mainly on VVV $Ks$-band observations,and its magnitude was not reported in the VISTA Science Archive \citep[VSA;][]{cros12}. The $Ks$-band magnitude (18.162) of VVV~1256$-$62A in Table~\ref{tprop} is from the 2010 image. 

The optical and NIR photometric SEDs of VVV~1256$-$62A and B and the X-shooter spectrum of VVV~1256$-$62B are all shown in Figure~\ref{fspec}. The WD primary is brighter in the optical bands ($gri$), while the L subdwarf companion is brighter in the NIR bands ($ZYJHKs$). The {\sl Gaia} $G$ and $G_{\rm RP}$ photometry of VVV~1256$-$62A is slightly fainter than the DECaPS SED profile, which may be due to
blending with background sources or unresolved spectral features present in the broad {\sl Gaia} filters.
The VVV $Y$- and $J$-band magnitudes of VVV~1256$-$62A measured in 2010 are slightly brighter than those in 2015, which is likely due to blending with nearby objects. The 2015 VVV SED profile is more consistent with the DECaPS SED profile, and the combined profile indicates that VVV~1256$-$62A is a cool WD; indeed, the faintest in $M_G$ in our sample (see Fig. \ref{fhrd}). 

\subsection{GMOS spectroscopy of VVV~1256$-$62A}
VVV~1256$-$62A was observed as part of a programme GS-2021A-FT-108 (PI: Schneider) with the Gemini Multi-Object Spectrograph \citep[GMOS;][]{gmos} on the Gemini South telescope. We used the longslit mode with the R400 grating and a central wavelength of 764~nm and the 1 arcsec slit with 2$\times$2 binning. We obtained four spectra, one on 2021 May 09 and three on 2021 May 11, each with an exposure time of 900~s. 

\begin{figure*}
	\includegraphics[width=\textwidth]{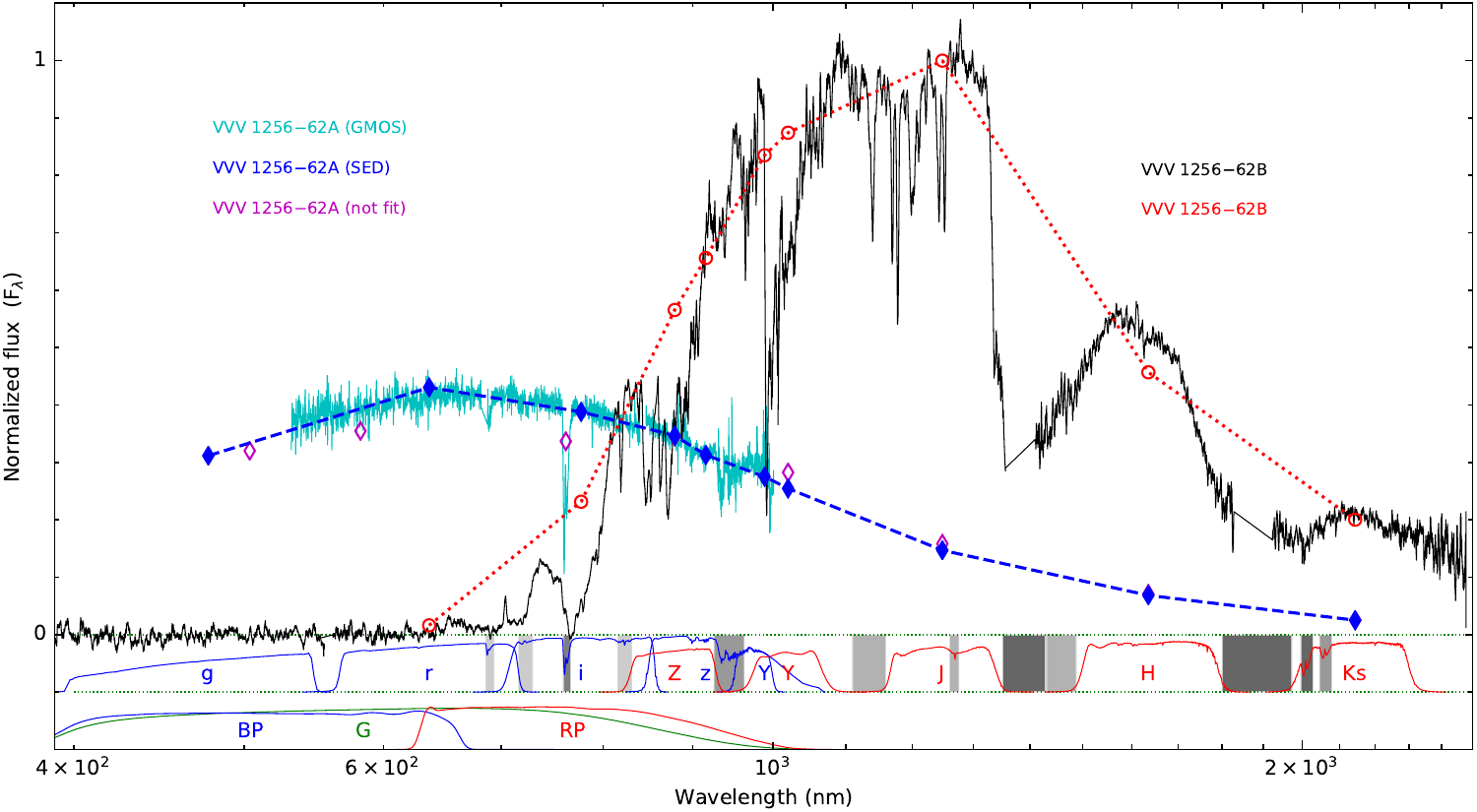}
    \caption{The X-shooter optical to NIR spectrum of VVV~1256$-$62B \citepalias{prime5} normalized in the 1100--1300~nm region. DECaPS \citep{decaps1} and VVV (2015 $ZYJH$ and 2010 $Ks$) photometry of VVV~1256$-$62A (filled diamonds) and B (open circles) are plotted and re-scaled by the same factor. GMOS optical spectrum of VVV~1256$-$62A is scaled to its SED. Note telluric absorptions (indicated with grey bands at the bottom) are corrected for the X-shooter spectrum, and not corrected for the GMOS spectrum. Open diamonds are {\sl Gaia} and VVV (2010) photometry of VVV~1256$-$62A. 
    DECam, VISTA, and {\sl Gaia} filter profiles are plotted at the bottom, with filter names labelled at their effective wavelength.}
    \label{fspec}
\end{figure*}

The data were reduced using the Gemini package \textsc{dragons} \citep{dragons} and the GMOS specific reduction routines for spectroscopy that perform the bias, flatfield and bad pixel corrections as well as determining the wavelength solution and response function before reducing the science frames and producing a 1D spectrum. The combined spectrum of VVV~1256$-$62A is plotted in Fig. \ref{fspec}.

\begin{figure*}
    	\includegraphics[width=0.495\textwidth]{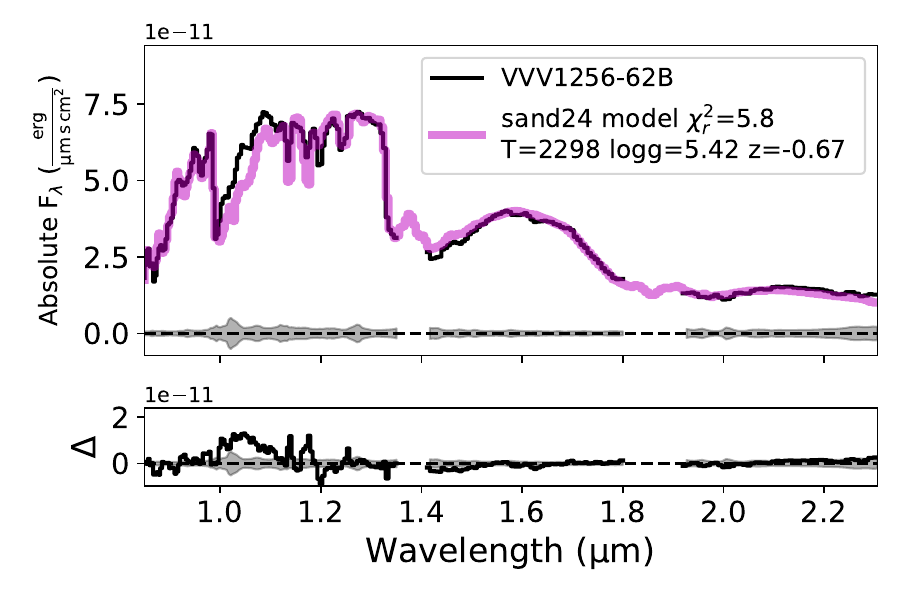} 
    \includegraphics[width=0.495\textwidth]{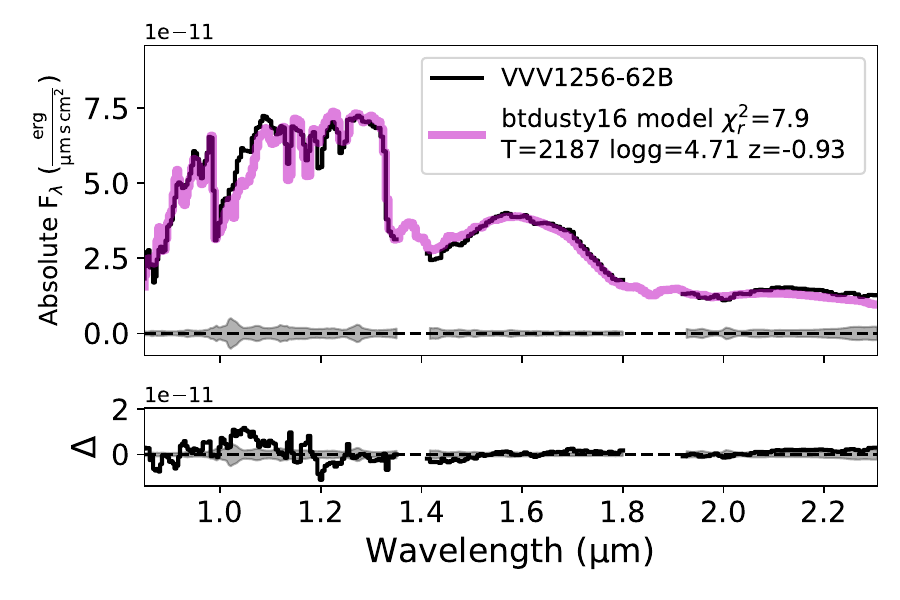} 
    \caption{(Top panel) The smoothed X-shooter spectrum of VVV~1256$-$62B calibrated to absolute flux densities compared to the best-fitting models from SAND (top) and BT-Dusty (bottom). 
    Each panel lists the $T_{\rm eff}$, $\log{g}$, and [M/H] of the best fit interpolated model, and weighted means and uncertainties are listed in Table~\ref{tab:modelfits}.
    The bottom panels display the difference between observed and computed spectra (black line) compared to the spectral uncertainty (grey band).}
    \label{fspecf}
\end{figure*}

\begin{figure*}
	\includegraphics[width=0.95\textwidth]{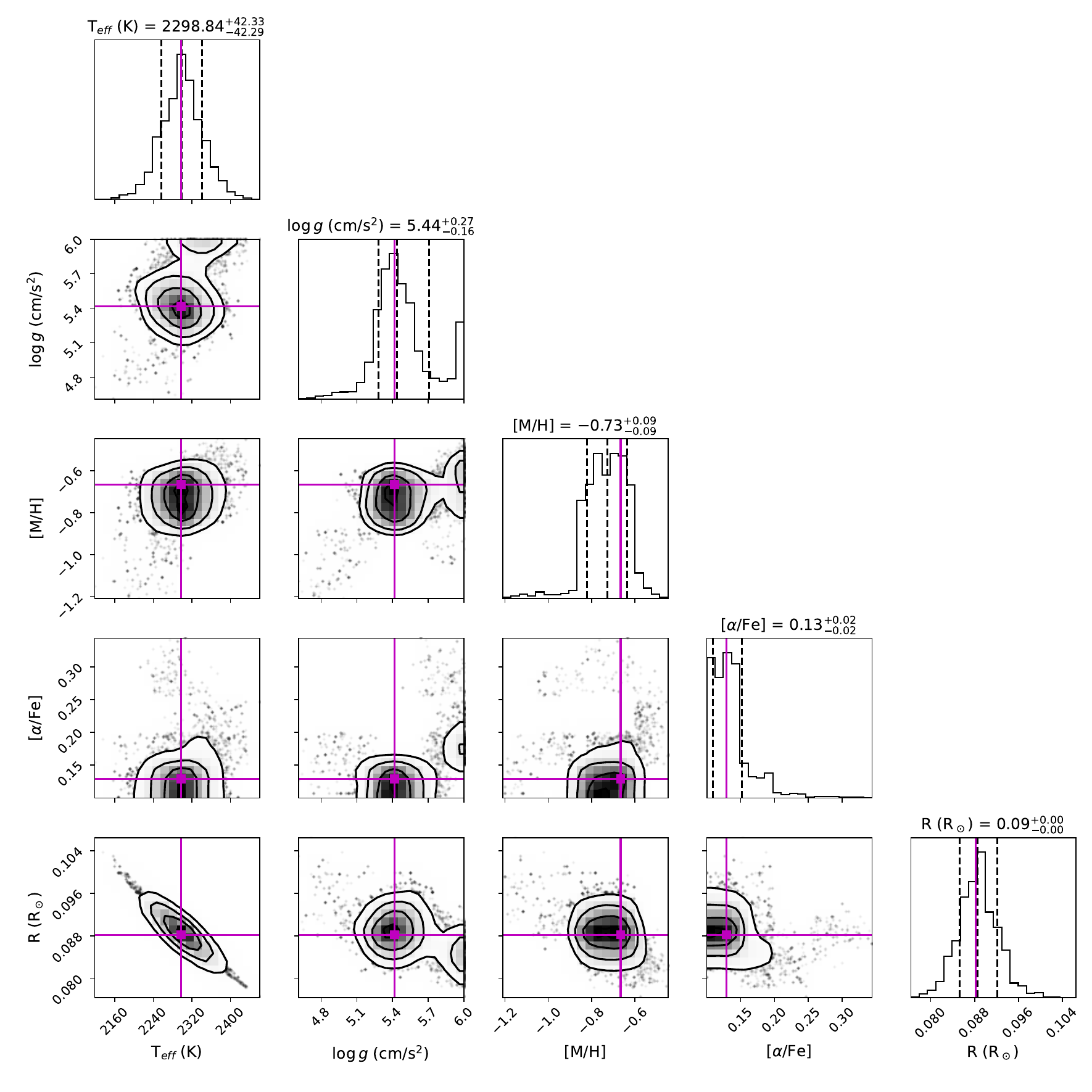}
    \caption{$T_{\rm eff}$, log$g$, [M/H], [$\alpha$/Fe], and radius distributions for the SAND MCMC model fits for VVV~1256$-$62B. Plots along the diagonal axis show the marginalized posterior distributions for each parameter, with 16 percent, 50 percent, and 84 percent quantiles indicated as vertical dashed lines. The remaining contour plots display two-dimensional distributions of parameter pairs in the posterior solutions, highlighting parameter correlations. 
    The lines and filled circles indicate the parameters of the single best-fitting model.}
    \label{ffitp}
\end{figure*}

\begin{figure*}
    \includegraphics[width=0.33\linewidth]{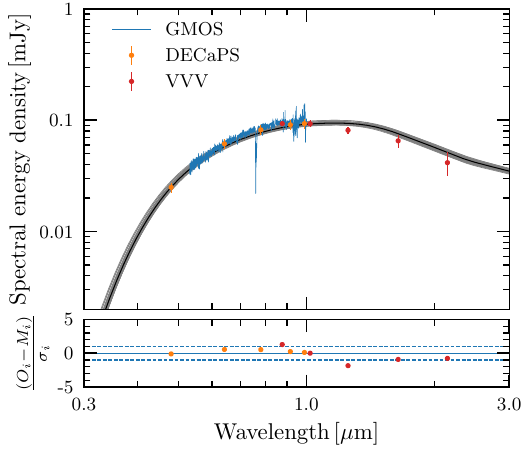}
    \includegraphics[width=0.33\linewidth]{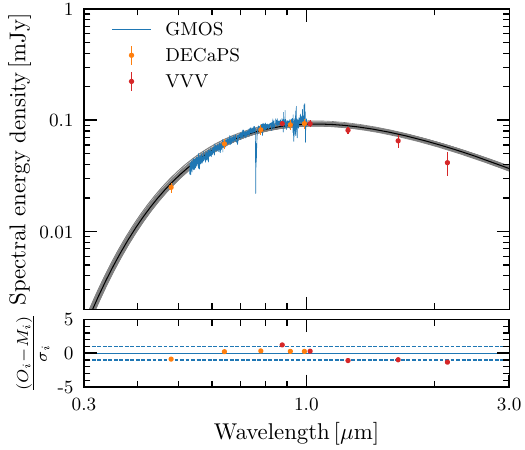} 
        \includegraphics[width=0.33\linewidth]{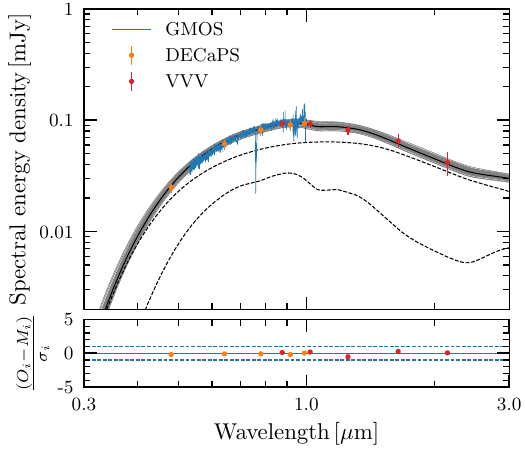}
    \caption{Best-fitting  models reproducing the DECaPS and the VVV photometry of VVV~1256$-$62A. The GMOS spectrum of this source is also plotted for comparison. Synthetic models (black curves) and fifty random draws corresponding to the estimated uncertainties (grey curves) are shown for hydrogen-dominated (left) and helium-dominated (middle) atmospheres. 
    The right panel shows the best-fitting model for an unresolved double white dwarf system, obtained by using two synthetic spectra of hydrogen-dominated atmospheres.
    The bottom sub-panels display the residuals between observed and synthetic photometry.}
    \label{fmodel1}
\end{figure*}

\begin{table*}
 \centering
  \caption[]{Physical parameters of VVV~1256$-$62B from atmosphere model fits.}
\label{tab:modelfits}
  \begin{tabular}{l c c c c c c}
\hline
Model & $T_{\rm eff}$ & $\log{g}$ & [M/H] & [$\alpha$/H] or C/O & Radius & $\chi^2_r$ \\
 & (K) & (cm~s$^{-2}$) &  &  &  (R$_\odot$) &  (Rel.~Prob.)$^a$ \\
\hline
SAND & 2298$^{+45}_{-43}$ & 5.43$^{+0.26}_{-0.15}$ & $-$0.72$^{+0.08}_{-0.10}$ & +0.13$^{+0.02}_{-0.02}$   & 0.089$^{+0.004}_{-0.003}$ &  5.8 (1.0) \\
BT-Dusty & 2220$^{+42}_{-30}$ & 5.1$^{+0.3}_{-0.4}$ & $-$0.87$^{+0.08}_{-0.08}$ & +0.03$^{+0.06}_{-0.03}$ &  0.096$^{+0.003}_{-0.004}$ & 7.9 (0.12) \\
Elf Owl & 2245$^{+83}_{-67}$ & 4.8$^{+0.4}_{-0.2}$ & $-$0.85$^{+0.20}_{-0.15}$ & C/O = 0.62$^{+0.31}_{-0.12}$ &  0.094$^{+0.006}_{-0.007}$ & 17 (10$^{-5}$) \\
Burrows06 & 2001$^{+167}_{-41}$ & 5.42$^{+0.09}_{-0.21}$ & $<-$0.50 & --- & 0.112$^{+0.005}_{-0.016}$ & 24 (10$^{-8}$)  \\
Bobcat & 2345$^{+46}_{-60}$ & 5.37$^{+0.13}_{-0.11}$ & $-$0.31$^{+0.25}_{-0.18}$ & C/O = 1.0$^{+0.5}_{-0.4}$   & 0.087$^{+0.005}_{-0.003}$ & 26 (10$^{-9}$) \\
Drift & 2440$^{+112}_{-111}$ & 5.29$^{+0.19}_{-0.22}$ & $-$0.54$^{+0.13}_{-0.06}$ &  --- & 0.078$^{+0.008}_{-0.006}$ & 28 (10$^{-10}$) \\
\hline
\end{tabular}
\begin{list}{}{}
\item[]$^a$Relative probability among the best-fit models computed as $-\ln{P} = \chi^2_r - {\rm MIN}(\{\chi^2_r\})$.
\end{list}
\end{table*}

\subsection{Physical properties of the L subdwarf companion}
\label{s1256b}
The physical properties of VVV~1256$-$62B were inferred by fitting its spectrum to six sets of low-temperature atmospheric models that include sub-solar metallicities\footnote{The LOWZ model set \citep{meis21} was excluded here as its model parameter range does not extend to $T_{\rm eff} >$ 1600~K.} and are contained in the SpeX Prism Libaries Analysis Toolkit \citep{burg17a}:
\citet{burr06}, 
Drift \citep{witt11}, 
BT-Dusty \citep{alla14}, 
Sonora Bobcat \citep{marl21} and Elf Owl \citep{mukh24}, and
Spectral ANalog of Dwarfs (SAND \citealt{Alvarado_2024}).
We combined the optical and NIR spectra of VVV~1256$-$62B, resampled to a resolution of $\lambda/\Delta\lambda$ = 250 over the wavelength range 0.85--2.4$\mu$m, and scaled to the absolute VVV $J$ magnitude based on the parallax of the WD primary.
The atmosphere models were also resampled. Because the atmosphere models are scaled to surface fluxes, the relative normalization between models and observed absolute fluxes ($\alpha$) provides an estimate of the source radius.\footnote{The conversion to radius is $R = \sqrt{\alpha}\times$10~pc = 2.255$\times$10$^{-9}\sqrt\alpha$~R$_\odot$.}
We used a $\chi^2$ goodness-of-fit statistic to identify the best-fitting model, first among the individual models in each of these grids, and then using this as an initial guess to a Metropolis-Hastings Monte Carlo Markov Chain (MCMC) fitting algorithm \citep{1953JChPh..21.1087M,HASTINGS01041970} the sampled the model parameter space around the best fitting parameters through linear interpolation of the log fluxes of the models (further description is provided in Burgasser et al., in prep.). 

Table~\ref{tab:modelfits} summarizes the resulting fit parameters, while Figures~\ref{fspecf} and~\ref{ffitp} display the best fitting spectrum and parameter distribution for the SAND and BT-Dusty models, which provided the best overall fits.
Both of these models provide a good overall match to the overall near-infrared spectral shape of VVV~1256$-$62B, and primarily deviate in reproducing spectral structure near the 1~$\mu$m FeH Wing-Ford band and Na-I and K~I line strengths in the 1.1--1.25~$\mu$m region. The mean parameters from these two model sets are consistent in $T_{\rm eff}$ and $\log{g}$, and notably yield statistically equivalent measures of sub-solar metallicity, [M/H] = $-$0.72$^{+0.08}_{-0.10}$ and $-$0.87$^{+0.08}_{-0.08}$ for SAND and BT-Dusty, respectively. There is also evidence of significant alpha abundance enhancement for the SAND models. The mean [Fe/H] of the SAND models is $-0.81\pm0.10$ according to $[M/H] \approx [Fe/H] + log_{10}(0.694 \times 10^{[\alpha/Fe]} +0.306)$ \citep{sala05}. The $\alpha$-enhancement looks marginal for the BT-Dusty models. This is because the [M/H] indicated in the BT-Dusty models are actually [Fe/H] accounted for $\alpha$-enhancement, [$\alpha$/Fe] = +0.2 is adopted for [Fe/H] = $-$0.5 and [$\alpha$/Fe] = +0.4 is adopted for [Fe/H] $\le$ –1.0 (also see \citealt[][hereafter, \citetalias{prime2}]{prime2}). 
Finally, both models yield flux-scaled radii that are equivalent and in line with expectations for evolved ELMS ($R \sim$ 0.09~R$_\odot$; \citealt{bara97}).

With $T_{\rm eff}$ = 2298$^{+45}_{-43}$~K and [Fe/H] = $-0.81\pm0.10$, VVV~1256$-$62B lies just above the stellar/substellar boundary in the $T_{\rm eff}$ versus [Fe/H] space (see fig. 9 in  \citetalias{prime2}). Both parameters are very sensitive to mass near the hydrogen burning limit, and we infer 
a mass of 0.082$\pm$0.001 M$_{\odot}$ based on the 10Gyr iso-mass contours \citep{burr98} of low mass objects in the $T_{\rm eff}$ vs [Fe/H] space.

\subsection{The age of cool white dwarf primary}
\label{s1256a}
The WD companion, VVV~1256$-$62A is located at the bottom right of the WD cooling sequence in the H-R diagram (Fig. \ref{fhrd}), consistent with being an old, low-mass WD \citep{raddi2022}. By fitting the {\em Gaia} EDR3 photometry of this WD, \citet{gent21} estimated a $T_{\rm eff} \approx 4560$\,K and $\log{g} \approx 7.9$, inferring a mass of about 0.52\,M$_\odot$. {\em Gaia} DR3 blue photometer (BP) and red photometer (RP) spectra \citep{BPRP_valid,BPRP_calib} are available for this source, but are rather noisy as expected for a faint object. 

We re-evaluated the physical parameters of VVV~1256$-$62A by analysing its DECaPS and VVV photometric SED that combines its DECaPS \citep{decaps1} and VVV photometry 
(Fig. \ref{fmodel1}). We favoured the DECaPS photometry over \emph{Gaia} DR3 photometry and BP/RP spectra due to the poorer quality of the latter. We employed state-of-the-art synthetic spectra of cool WDs with hydrogen- or helium-dominated atmospheres that include collision-induced absorption effects \citep{saumon2006,tremblay2011,tremblay2013,cukanovaite2021} to compute synthetic photometry in the observed band-passes. 
Our fitting routine minimizes the $\chi^2$ between the observed and synthetic photometry. 
We scaled the synthetic photometry to the mass-radius relations of cooling models for hydrogen-dominated atmospheres \citep{althaus2013,camisassa2016} and hydrogen-deficient atmospheres \citep{blouin2018,bedard2020}. We also considered a mass-radius relation for hydrogen-dominated WDs with metal-poor progenitors \citep[$Z = 0.001$;][]{serenelli2002,althaus2015}, which may be more appropriate given the sub-solar metallicity of the L subdwarf companion. The \emph{Gaia} parallax provides a distance prior, and at just 75.6\,pc from the Sun we expect interstellar extinction to be small ($A_v = 0.005$\,mag) based on distance-reddening relations \citep{lallement2019}.

The best-fits for our hydrogen-dominated and hydrogen-deficient models
(Fig. \ref{fmodel1})
are numerically consistent, with reduced $\chi^2$ values close to unity. The hydrogen-dominated atmosphere model may be more realistic given the apparent decline in the near-infrared photometry that is not correctly captured in the models due to current uncertainties in the treatment of the H$_2$--He collision-induced-absorption in cool WD atmospheres \citep{blouin2018}. 
Our inferred atmosphere parameters of $T_{\rm eff} = 4440 \pm 250$\,K and $\log{g}= 7.86 \pm 0.05$\,dex for the hydrogen-dominated model, or $T_{\rm eff} = 4550 \pm 250$\,K and $\log{g}= 7.88 \pm 0.05$\,dex for the hydrogen-deficient model, yield similar masses of $0.51\pm 0.03$\,M$_\odot$, and these results are compatible with those obtained by \citet{gent21}. 

For its present-day temperature and mass, we estimate a cooling age for the solar and sub-solar metallicity hydrogen-dominated evolutionary models of $6.25\,\pm 1.30$ and $8.90\pm1.35$\,Gyr, respectively. The hydrogen-deficient model yields a cooling age of $6.75 \pm 0.80$\,Gyr. However, there is a problem with this solution as the semi-empirical initial-to-final-mass relation \citep[IFMR;][]{catalan2008} applied to the present-day mass of VVV~1256$-$62A, $0.51\pm 0.03$\,M$_\odot$, yields a progenitor mass of 0.84\,M$_\odot$. The Main Sequence lifetime of such a low-mass progenitor, $\sim$15~Gyr, would result in a total age that exceeds the age of the Universe.

VVV~1256$-$62A is at the end of the WD cooling sequence where evolutionary tracks turn toward the blue. In this region, photometric estimates of WD masses are known to be systematically smaller with respect to the average mass of $\approx 0.6$\,M$_\odot$ of the broader WD population \citep[][and references therein]{obrien2024}, which is expected to be independent of temperature for a population of WDs evolving through single-star evolution \citep{tremblay2016}. One proposed hypothesis is that cool low-mass WDs could be unresolved binaries; however, such a situation would require an unrealistically high fraction of binaries at low temperatures \citep{obrien2024}. Another hypothesis, taken into account by \citet{obrien2024}, is that optical opacities in WD atmospheres are presently underestimated \citep[e.g. associated with the red wing of Ly$\alpha$;][]{caron2023}. These authors provide an \emph{ad-hoc} correction that preserves an average WD mass at cool temperatures \citep[eq. 1, 2, and 3;][]{obrien2024}. Applying this correction, we obtain a corrected mass of $0.62 \pm 0.04$\,M$_\odot$ which shifts the cooling ages to $8.0 \pm 1.9$ and $8.5 \pm 1.9$\,Gyr for the solar and sub-solar metallicity hydrogen-dominated models, and $7.0\pm 0.8$ Gyr for the hydrogen-deficient model.
More importantly, the corrected WD mass implies a progenitor mass of $1.9 \pm 0.4$\,M$_\odot$ which has a main-sequence lifetime of $2.0^{+1.8}_{-0.8}$\,Gyr based on the $\mathrm{[Fe/H]} =+0.06$\,dex evolutionary tracks by \citet{hidalgo2018}, where the uncertainties reflect 16 percent and 84 percent quantiles. These values yield a total age of $10^{+2.7}_{-2.1}$\,Gyr. 
For the hydrogen-deficient WD model, the total age is $\approx1$\,Gyr shorter.
For the sub-solar metallicity hydrogen-dominated model, the theoretical IFMR of \citet{romero2015} with $Z = 0.001$ yields a lower progenitor mass of $1.5 \pm 0.3$\,M$_{\odot}$, which in turn implies a progenitor age of $2.0^{+2.3}_{-0.8}$\,Gyr based on $\alpha$-enhanced [Fe/H]$= -0.81$\,dex evolutionary models \citep{pietrinferni2021}. These results yield a total age of $10.5^{+3.3}_{-2.1}$\,Gyr. 
Given that metallicity of the L subdwarf companion most supports the sub-solar metallicity hydrogen-dominated model, we adopt this last   total age estimate for the VVV~1256$-$62AB system (Table\,\ref{tprop}).

Despite the reliability of cooling processes in WDs, there are several sources of uncertainty that can further influence the inferred age of this system due to the complex processes occurring in cool atmospheres, most notably strong collision-induced absorption effects \citep{blouin2018}. Additionally, a cooling delay or even speed-up can occur depending on the onset of crystallization in WD cores \citep{bauer2020,camisassa2024}.
Nevertheless, our final age estimate is relatively insensitive to the specific white dwarf type (hydrogen-dominated versus hydrogen-deficient), and is compatible with the ages of Galactic thick disc and halo stars as indicated by the kinematics of this source (see Section \ref{ski}). 

For the sake of completeness, we also investigated the hypothesis that VVV~1256$-$62A could be an unresolved WD binary based on its position on the H-R diagram. We employed the same minimization routine using two hydrogen-dominated WD models. We obtained a slightly improved best fit with smaller residuals (Fig.\,\ref{fmodel1} right). 
The two WDs would have masses of $0.62\pm 0.03$ and $0.72 \pm 0.03$\,M$_\odot$, with cooling ages of $13.6\pm1.1$ and $9.8\pm1.4$ Gyr. The estimated total ages of $15.4^{+1.9}_{-1.4}$ and $10.5^{+1.9}_{-2.0}$ Gyr are discrepant by $\approx 5$\,Gyr, but compatible within $2\,\sigma$.
While this situation is implausible for a binary system whose components should have formed at the same time, the cooling age delays noted above and the relatively large uncertainties could resolve the incompatible age estimates in a binary. None the less, there is no clear evidence that VVV~1256$-$62A could be an unresolved double WD, and the small Renormalized Unit Weight Error (RUWE: 1.034) from {\sl Gaia} DR3 for this source further argues agains the binary hypothesis.

\begin{figure*}
	\includegraphics[width=\textwidth]{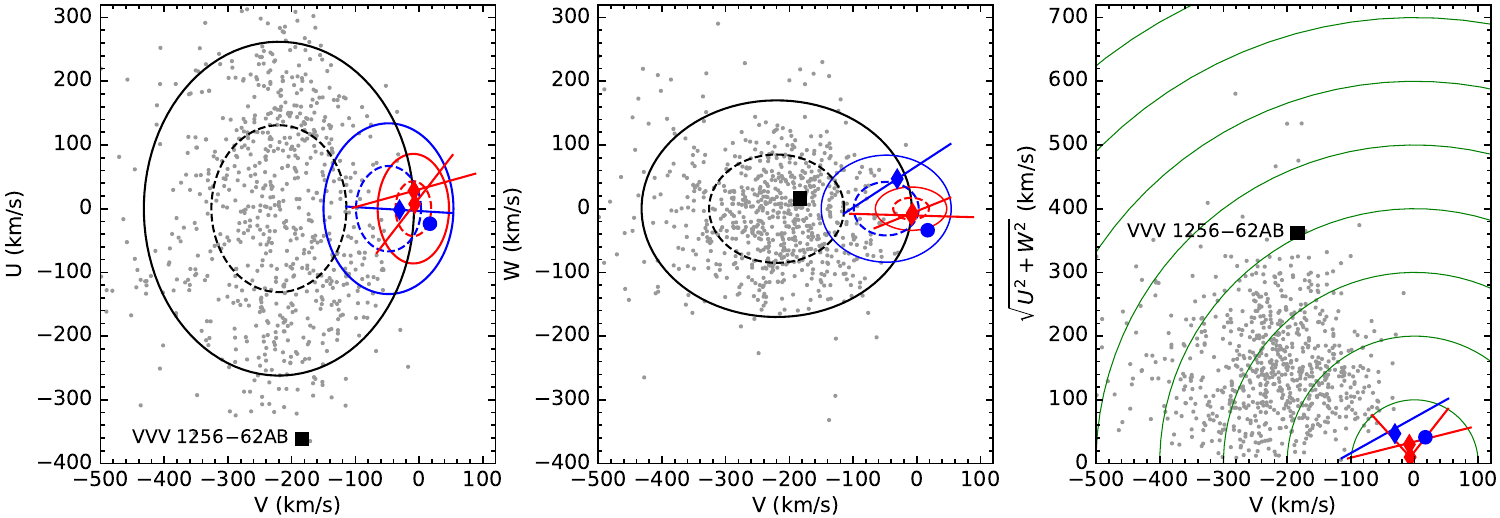}
    \caption{Space velocities of five WD + UCD wide binary systems compared to esdM and usdM subdwarfs \citep[grey dots;][]{zhan13}. The circles from right to left are 1$\sigma$ (dashed) and 2$\sigma$ (solid) velocity dispersions of the Galactic thin disc, thick disc, and halo, respectively \citep{redd06}. Three systems (diamonds) with error bars indicate uncertainties caused by RV variations for thin disc objects ($-$100, 0, 100 km s$^{-1}$; e.g. fig. 4 of \citealt{zhan13}). The error bars for SD1555 AB (filled circle) and VVV~1256$-$62AB (filled square) are smaller than the symbol size.}
    \label{fuvw}
\end{figure*}

\begin{figure*}
    \centering
    \includegraphics[width=1.03\textwidth]{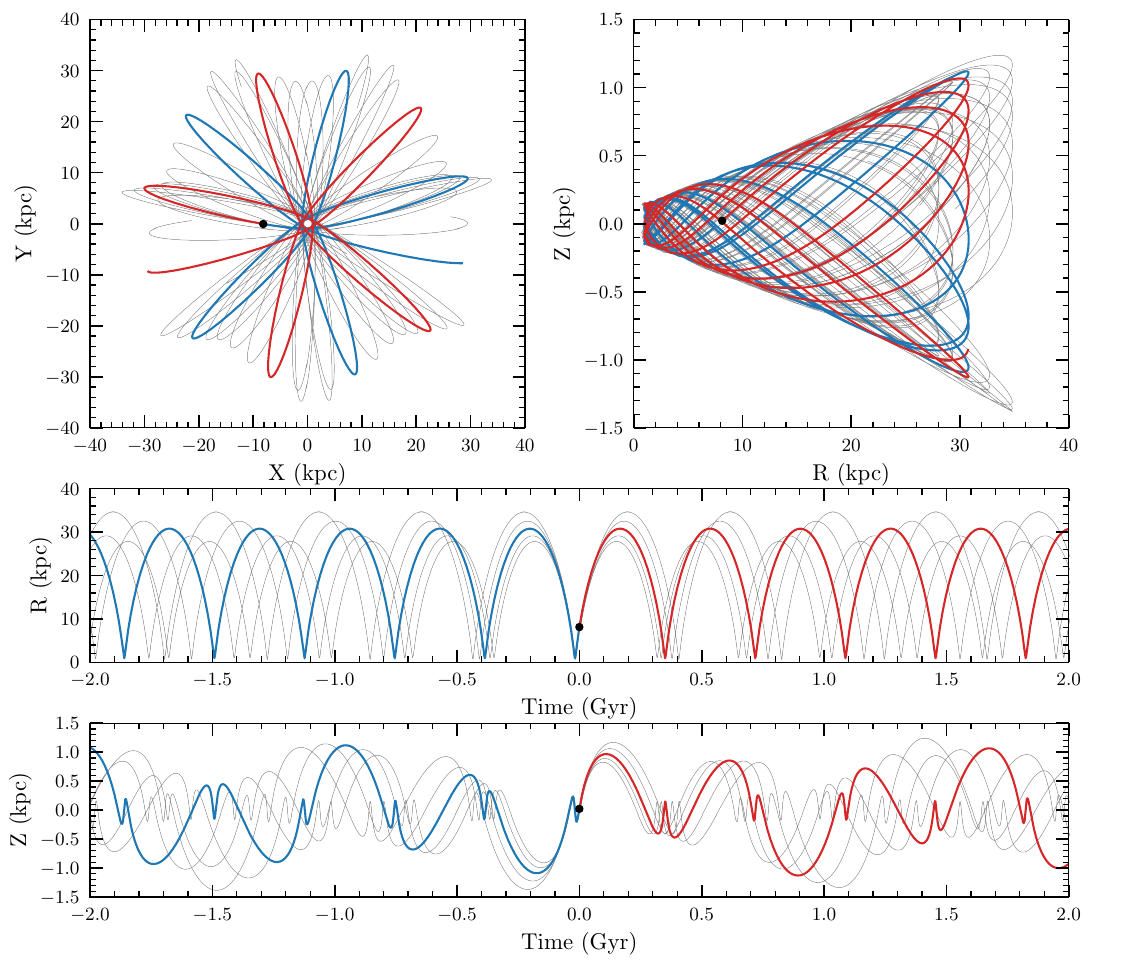} 
\caption{Integrated Galactic orbit of VVV~1256$-$62AB over an interval of 4\,Gyr. The thick blue and red curves, corresponding to past and future motion, represent the average orbit based on the {\em Gaia} measurements of VVV~1256$-$62A and the RV of VVV~1256$-$62B.
The grey curves represent four orbits that account for the $\pm$1$\sigma$ and $\pm$2$\sigma$ variation in the orbital trajectory due to propagation of measurement uncertainties. 
(Top left): the orbit in Galactic Cartesian coordinates. The Sun is placed at $X, Y, Z$ = ($-$8.178, 0, 0.021) in kpc \citep{grav19,bennet2019}, the Galactic rotation is clockwise. The black dot at ($-$8.136, $-$0.063, 0.022) in kpc representing the current location of the system.
(Top right:) the orbit in Galactic cylindrical coordinates. 
(Middle:) the Galactocentric radius as a function of time.  
(Bottom:) the vertical displacement as a function of time.}
    \label{fig:orbits}
\end{figure*}

\subsection{Kinematics}
\label{ski}
We measured the heliocentric RV of VVV~1256$-$62B from its X-shooter spectrum (Fig. \ref{fspec}) using the cross-correlation technique  \citep[e.g.][]{Galvez_2002}. The spectra of the target were cross-correlated using the routine $fxcor$ in the Image Reduction and Analysis Facility \citep[IRAF, e.g.][]{tody1993}, against spectra of a radial velocity standard, DENIS-P J144137.3$-$094559 \citep[DE1441;][]{mart99}, an L0.5 with known radial velocity ($-27.9 \pm 1.2$ km s$^{-1}$ \citep{bail04}. We derived the radial velocity from the position of peak of the cross-correlation function (CCF) and calculated uncertainties based on the fitted peak height and the antisymmetric noise as described by \citet{tonryanddavis1979}. Areas affected by prominent telluric lines were excluded when determining the mean velocity.

Using the RV from the L subdwarf and the {\sl Gaia} DR3 astrometry from the WD, we
calculated the Galactic velocities\footnote{These velocity components are defined such that 
$U$ is positive toward the Galactic centre,
$V$ is positive in the direction of Galactic rotation,
and $W$ is positive toward the North Galactic Pole.}
($UVW$) of VVV~1256$-$62AB using the BANYAN $\Sigma$ interface \citep{gagn18}. 
Fig. \ref{fuvw} compares these velocities to the other WD + M (sub)dwarf companions in our sample and esdM and usdM subdwarfs compiled in \citet{zhan13}.
The velocity components of this system are extreme even among M subdwarfs, with very large tangential velocity ($403\pm10$ km s$^{-1}$)
and total space velocity ($406\pm10$ km s$^{-1}$), confirming its halo membership. 

We integrated the Galactic orbit of VVV~1256$-$62AB assuming a static three-component potential that is representative of the Milky Way's baryonic and dark matter components (the \verb|MWPotential2014| of the \verb|galpy| module for \verb|python|; \citet{bovy2015} for details). We accounted for the uncertainties of the velocity components be examining both the median trajectory and $\pm$1$\sigma$ variants. A visual representation of the Galactic orbit and the time evolution of the cylindrical radial and vertical coordinates are shown in Fig.,\ref{fig:orbits}. As reported in \citetalias{prime5}, this orbit is highly eccentric ($e \approx 0.9$), prograde ($L_Z \approx 600$\,kpc\,km/s), and moving radially away from the Galactic centre. 
VVV~1256$-$62AB passes close to the inner radius ($\sim$1 kpc) of the Milky Way, making dynamical interactions with the Galactic bar likely. VVV~1256$-$62AB also spends about 2/3 of its time at $R >$ 20 kpc, a region dominated by outer halo population \citep{caro07}.
Its low-inclination orbit is not unexpected for a source within a few hundred parsecs away from the Sun, and is likely a selection effect of its proximity.
Overall, this source exhibits kinematics and orbital properties consistent with halo membership, as has been found for other WDs in the solar neighbourhood \citep{zubi24}.

\subsection{Stability of the wide binary}
The lifetime of a wide binary, which has resisted disruption from perturbations due to passing field stars, is proportional to the average relative velocity ($V_{\rm rel}$) between the binary system and perturbers and inversely proportional to the number density ($n^*$) of field stars \citep[equation 28,]{wein87}. VVV~1256$-$62AB passes through the Galactic bulge, coming as close as $\sim$1 kpc to the Galactic centre, where the stellar number density is approximately 10 times higher \citep{vale16}.
These higher densities should dissolve wide systems; however, $V_{\rm rel}$ is also about 10 times higher for VVV~1256$-$62AB compared to other stars in the solar neighbourhood. Thus, the binary lifetime of this system is comparable to similarly-separated wide binaries near the Sun, and is thus expected to be stable even during its passage through the denser regions of the Milky Way.

One approach to quantifying the stability of VVV~1256$-$62AB is through the tidal or Jacobi radius ($r_{\rm J}$), the boundary at which the Galactic tidal field becomes stronger than the gravitational attraction between the components of a wide binary. This scale is used to separate stable and unstable wide binaries. The Jacobi radius in the solar neighbourhood can be quantified as (equation (43) in \citealt{jian10}): 
\begin{equation}
   r_{\rm J}=1.70 {\rm pc} \left(\frac{M_1+M_2}{\rm 2M_{\sun}}\right)^{1/3}
	\label{erj}
\end{equation}
For VVV~1256$-$62AB, $r_{\rm J}$ = 1.2 pc or $2.5 \times 10^{5}$ au.
Hence, this system's projected separation of 1375$^{+35}_{-33}$ au at a distance of 75.6 pc corresponds to $5.56 \times 10^{3} r_{\rm J}$, well within the limit at which the  
Galactic tidal field is unable to disrupt the system.
It is therefore logical that VVV~1256$-$62AB has survived as a bound wide binary over its 10~Gyr lifetime.

\section{Conclusions}
\label{scon}
We have reported the discovery of five widely separated white dwarf + ELMS binary systems in the {\sl Gaia} Catalogue of Nearby Stars \citep{smar21}. Three of these systems are composed of WD + M dwarf pairs, one (SD1555+31AB) is composed of a WD + sdM9.5 subdwarf pair, and one (VVV~1256$-$62AB) is composed of a WD + L subdwarf pair. All five systems are confirmed by the common PM and equal parallactic distance, and span projected separations of 650-6000~au.

VVV~1256$-$62B is the first L subdwarf identified as a companion to a WD, and hence the first L dwarf age benchmark at sub-solar metallicities. We confirmed the sub-solar metallicity of this source through atmosphere model fits, finding [M/H] = $-0.72^{+0.08}_{-0.10}$ ([Fe/H] $-0.81\pm0.10$), in agreement with its metallicity classification and with prior analysis in \citetalias{prime5}. We were also able to determine its radial velocity, which combined with the precise astrometry of VVV~1256$-$62A from {\sl Gaia} DR3 yields space velocities that confirm kinematic membership in the Galactic halo.

VVV~1256$-$62A is a cool, likely hydrogen-dominated WD whose total age (cooling and progenitor) of 10.5$^{+3.3}_{-2.1}$~Gyr is also consistent with halo membership. Combined, these measurements yield a consistent picture of the VVV~1256$-$62AB system as a metal-poor, wide halo binary and a unique benchmark for testing both UCD and WD atmosphere and evolutionary models.
We also validate VVV~1256$-$62AB is a stable bound system despite a Galactic orbit that takes it from within the Galactic bulge to 25-35~kpc from the Galactic centre.

\section*{Acknowledgements}
Based on observations collected at the European Organisation for Astronomical Research in the Southern Hemisphere under ESO programme 0101.C-0626. 
Based on observations (Program ID: GS-2021A-FT-108; PI: Schneider) obtained at the international Gemini Observatory, a program of NSF NOIRLab (processed using the Gemini IRAF package [DRAGONS (Data Reduction for Astronomy from Gemini Observatory North and South)]), which is managed by the Association of Universities for Research in Astronomy (AURA) under a cooperative agreement with the U.S. National Science Foundation on behalf of the Gemini Observatory partnership: the U.S. National Science Foundation (United States), National Research Council (Canada), Agencia Nacional de Investigaci\'{o}n y Desarrollo (Chile), Ministerio de Ciencia, Tecnolog\'{i}a e Innovaci\'{o}n (Argentina), Minist\'{e}rio da Ci\^{e}ncia, Tecnologia, Inova\c{c}\~{o}es e Comunica\c{c}\~{o}es (Brazil), and Korea Astronomy and Space Science Institute (Republic of Korea).
This work has made use of data from the European Space Agency (ESA) mission
{\it Gaia} (\url{https://www.cosmos.esa.int/gaia}), processed by the {\it Gaia} Data Processing and Analysis Consortium (DPAC,
\url{https://www.cosmos.esa.int/web/gaia/dpac/consortium}). Funding for the DPAC has been provided by national institutions, in particular the institutions participating in the {\it Gaia} Multilateral Agreement.
DECaPS is based on observations at Cerro Tololo Inter-American Observatory, National Optical Astronomy Observatory (NOAO Prop. ID: 2014A-0429, 2016A-0327, and 2016B-0279; PI: Finkbeiner), which is operated by the Association of Universities for Research in Astronomy (AURA) under a cooperative agreement with the National Science Foundation.
Based on observations obtained as part of the VIKING survey from VISTA at the ESO Paranal Observatory, programme ID 179.A-2004 and 179.B-2002. Data processing has been contributed by the VISTA Data Flow System at CASU, Cambridge and WFAU, Edinburgh. The VISTA Data Flow System pipeline processing and science archive are described in  \citet{irwi04}, \citet{hamb08} and \citet{cros12}. 
This publication makes use of data products from the {\sl Wide-field Infrared Survey Explorer}, which is a joint project of the University of California, Los Angeles, and the Jet Propulsion Laboratory/California Institute of Technology, funded by the National Aeronautics and Space Administration.
Funding for the SDSS and SDSS-II has been provided by the Alfred P. Sloan Foundation, the Participating Institutions, the National Science Foundation, the U.S. Department of Energy, the National Aeronautics and Space Administration, the Japanese Monbukagakusho, the Max Planck Society, and the Higher Education Funding Council for England. The SDSS Web Site is \url{http://www.sdss.org/}. Funding for SDSS-III has been provided by the Alfred P. Sloan Foundation, the Participating Institutions, the National Science Foundation, and the U.S. Department of Energy Office of Science. The SDSS-III web site is \url{http://www.sdss3.org/}. This publication makes use of VOSA, developed under the Spanish Virtual Observatory project supported from the Spanish MICINN through grant AyA2008-02156. 
This publication makes use of the SpeX Prism Libraries Analysis Toolkit, maintained by Adam Burgasser at \url{https://github.com/aburgasser/splat}.

ZHZ acknowledges the supports from the Fundamental Research Funds for the Central Universities in China (14380034), the fundamental research programme of Jiangsu Province (BK20211143), the Program for Innovative Talents, Entrepreneur in Jiangsu (JSSCTD202139), and the science research grants from the China Manned Space Project with NO. CMS-CSST-2021-A08. RR acknowledges support from Grant RYC2021-030837-I funded by MCIN/AEI/10.13039/501100011033 and by “European Union NextGeneration EU/PRTR”. This work was partially supported by the AGAUR/Generalitat de Catalunya grant SGR-386/2021 and by the Spanish MINECO grant PID2020-117252GB-I00. M.C.G.O. acknowledges financial support from the Agencia Estatal de Investigación (AEI/10.13039/501100011033) of the Ministerio de Ciencia e Innovación and the ERDF “A way of making Europe” through project PID2022-137241NB-C42. BG is supported by the Polish National Science Center (NCN) under SONATA grant No. 2021/43/D/ST9/0194. YVP’s investigations were carried out under the MSCA4Ukraine program, project number 1.4-UKR-1233448-MSCA4Ukraine, which was funded by the European Commission.
The authors would like to thank Adam Schneider for providing the optical spectrum of VVV~1256$-$62A. The authors would like to thank the anonymous reviewer for his/her valuable comments. 


\section*{Data Availability}
The X-shooter spectrum of VVV 1256$-$62B underlying this article is available in \citetalias{prime5} at \url{https://dx.doi.org/10.1093/mnras/stz659}. The optical spectrum of SD1555+31B is available in the SDSS database at \url{https://skyserver.sdss.org/}. The GMOS optical spectrum of VVV 1256$-$62A and a video of VVV 1256$-$62AB orbiting the Milky Way are available online.



\bibliographystyle{mnras}
\bibliography{example} 

\begin{thebibliography}{}
\makeatletter
\relax
\def\mn@urlcharsother{\let\do\@makeother \do\$\do\&\do\#\do\^\do\_\do\%\do\~}
\def\mn@doi{\begingroup\mn@urlcharsother \@ifnextchar [ {\mn@doi@} {\mn@doi@[]}}
\def\mn@doi@[#1]#2{\def\@tempa{#1}\ifx\@tempa\@empty \href {http://dx.doi.org/#2} {doi:#2}\else \href {http://dx.doi.org/#2} {#1}\fi \endgroup}
\def\mn@eprint#1#2{\mn@eprint@#1:#2::\@nil}
\def\mn@eprint@arXiv#1{\href {http://arxiv.org/abs/#1} {{\tt arXiv:#1}}}
\def\mn@eprint@dblp#1{\href {http://dblp.uni-trier.de/rec/bibtex/#1.xml} {dblp:#1}}
\def\mn@eprint@#1:#2:#3:#4\@nil{\def\@tempa {#1}\def\@tempb {#2}\def\@tempc {#3}\ifx \@tempc \@empty \let \@tempc \@tempb \let \@tempb \@tempa \fi \ifx \@tempb \@empty \def\@tempb {arXiv}\fi \@ifundefined {mn@eprint@\@tempb}{\@tempb:\@tempc}{\expandafter \expandafter \csname mn@eprint@\@tempb\endcsname \expandafter{\@tempc}}}

\bibitem[\protect\citeauthoryear{{Allard}}{{Allard}}{2014}]{alla14}
{Allard} F.,  2014, in {Booth} M.,  {Matthews} B.~C.,   {Graham} J.~R.,  eds, ~ Vol. 299, Exploring the Formation and Evolution of Planetary Systems. pp 271--272, \mn@doi{10.1017/S1743921313008545}

\bibitem[\protect\citeauthoryear{{Althaus}, {C{\'o}rsico}, {Isern}  \& {Garc{\'\i}a-Berro}}{{Althaus} et~al.}{2010}]{althaus2010}
{Althaus} L.~G.,  {C{\'o}rsico} A.~H.,  {Isern} J.,   {Garc{\'\i}a-Berro} E.,  2010, \mn@doi [\aapr] {10.1007/s00159-010-0033-1}, \href {https://ui.adsabs.harvard.edu/abs/2010A&ARv..18..471A} {18, 471}

\bibitem[\protect\citeauthoryear{{Althaus}, {Miller Bertolami}  \& {C{\'o}rsico}}{{Althaus} et~al.}{2013}]{althaus2013}
{Althaus} L.~G.,  {Miller Bertolami} M.~M.,   {C{\'o}rsico} A.~H.,  2013, \mn@doi [\aap] {10.1051/0004-6361/201321868}, \href {https://ui.adsabs.harvard.edu/abs/2013A&A...557A..19A} {557, A19}

\bibitem[\protect\citeauthoryear{{Althaus}, {Camisassa}, {Miller Bertolami}, {C{\'o}rsico}  \& {Garc{\'\i}a-Berro}}{{Althaus} et~al.}{2015}]{althaus2015}
{Althaus} L.~G.,  {Camisassa} M.~E.,  {Miller Bertolami} M.~M.,  {C{\'o}rsico} A.~H.,   {Garc{\'\i}a-Berro} E.,  2015, \mn@doi [\aap] {10.1051/0004-6361/201424922}, \href {https://ui.adsabs.harvard.edu/abs/2015A&A...576A...9A} {576, A9}

\bibitem[\protect\citeauthoryear{Alvarado, Gerasimov, Burgasser, Brooks, Aganze  \& Theissen}{Alvarado et~al.}{2024}]{Alvarado_2024}
Alvarado E.,  Gerasimov R.,  Burgasser A.~J.,  Brooks H.,  Aganze C.,   Theissen C.~A.,  2024, \mn@doi [Research Notes of the AAS] {10.3847/2515-5172/ad4bd7}, 8, 134

\bibitem[\protect\citeauthoryear{{Bailer-Jones}}{{Bailer-Jones}}{2004}]{bail04}
{Bailer-Jones} C.~A.~L.,  2004, \mn@doi [\aap] {10.1051/0004-6361:20040965}, \href {http://adsabs.harvard.edu/abs/2004A%26A...419..703B} {419, 703}

\bibitem[\protect\citeauthoryear{{Baraffe}, {Chabrier}, {Allard}  \& {Hauschildt}}{{Baraffe} et~al.}{1997}]{bara97}
{Baraffe} I.,  {Chabrier} G.,  {Allard} F.,   {Hauschildt} P.~H.,  1997, \mn@doi [\aap] {10.48550/arXiv.astro-ph/9704144}, \href {https://ui.adsabs.harvard.edu/abs/1997A&A...327.1054B} {327, 1054}

\bibitem[\protect\citeauthoryear{{Bauer}, {Schwab}, {Bildsten}  \& {Cheng}}{{Bauer} et~al.}{2020}]{bauer2020}
{Bauer} E.~B.,  {Schwab} J.,  {Bildsten} L.,   {Cheng} S.,  2020, \mn@doi [\apj] {10.3847/1538-4357/abb5a5}, \href {https://ui.adsabs.harvard.edu/abs/2020ApJ...902...93B} {902, 93}

\bibitem[\protect\citeauthoryear{{Becklin} \& {Zuckerman}}{{Becklin} \& {Zuckerman}}{1988}]{beck88}
{Becklin} E.~E.,  {Zuckerman} B.,  1988, \mn@doi [\nat] {10.1038/336656a0}, \href {https://ui.adsabs.harvard.edu/abs/1988Natur.336..656B} {336, 656}

\bibitem[\protect\citeauthoryear{{B{\'e}dard}, {Bergeron}, {Brassard}  \& {Fontaine}}{{B{\'e}dard} et~al.}{2020}]{bedard2020}
{B{\'e}dard} A.,  {Bergeron} P.,  {Brassard} P.,   {Fontaine} G.,  2020, \mn@doi [\apj] {10.3847/1538-4357/abafbe}, \href {https://ui.adsabs.harvard.edu/abs/2020ApJ...901...93B} {901, 93}

\bibitem[\protect\citeauthoryear{{Bennett} \& {Bovy}}{{Bennett} \& {Bovy}}{2019}]{bennet2019}
{Bennett} M.,  {Bovy} J.,  2019, \mn@doi [\mnras] {10.1093/mnras/sty2813}, \href {https://ui.adsabs.harvard.edu/abs/2019MNRAS.482.1417B} {482, 1417}

\bibitem[\protect\citeauthoryear{{Best} et~al.,}{{Best} et~al.}{2018}]{best18}
{Best} W. M.~J.,  et~al., 2018, \mn@doi [\apjs] {10.3847/1538-4365/aa9982}, \href {https://ui.adsabs.harvard.edu/abs/2018ApJS..234....1B} {234, 1}

\bibitem[\protect\citeauthoryear{{Blouin}, {Dufour}  \& {Allard}}{{Blouin} et~al.}{2018}]{blouin2018}
{Blouin} S.,  {Dufour} P.,   {Allard} N.~F.,  2018, \mn@doi [\apj] {10.3847/1538-4357/aad4a9}, \href {https://ui.adsabs.harvard.edu/abs/2018ApJ...863..184B} {863, 184}

\bibitem[\protect\citeauthoryear{{Bochanski}, {West}, {Hawley}  \& {Covey}}{{Bochanski} et~al.}{2007}]{boch07}
{Bochanski} J.~J.,  {West} A.~A.,  {Hawley} S.~L.,   {Covey} K.~R.,  2007, \mn@doi [\aj] {10.1086/510240}, \href {https://ui.adsabs.harvard.edu/abs/2007AJ....133..531B} {133, 531}

\bibitem[\protect\citeauthoryear{{Bovy}}{{Bovy}}{2015}]{bovy2015}
{Bovy} J.,  2015, \mn@doi [\apjs] {10.1088/0067-0049/216/2/29}, \href {https://ui.adsabs.harvard.edu/abs/2015ApJS..216...29B} {216, 29}

\bibitem[\protect\citeauthoryear{{Burgasser}}{{Burgasser}}{2004}]{burg04}
{Burgasser} A.~J.,  2004, \mn@doi [\apjl] {10.1086/425418}, \href {http://ads.ari.uni-heidelberg.de/abs/2004ApJ...614L..73B} {614, L73}

\bibitem[\protect\citeauthoryear{{Burgasser} \& {Mamajek}}{{Burgasser} \& {Mamajek}}{2017}]{burg17}
{Burgasser} A.~J.,  {Mamajek} E.~E.,  2017, \mn@doi [\apj] {10.3847/1538-4357/aa7fea}, \href {http://adsabs.harvard.edu/abs/2017ApJ...845..110B} {845, 110}

\bibitem[\protect\citeauthoryear{{Burgasser} \& {Splat Development Team}}{{Burgasser} \& {Splat Development Team}}{2017}]{burg17a}
{Burgasser} A.~J.,  {Splat Development Team} 2017, in Astronomical Society of India Conference Series. pp 7--12 (\mn@eprint {arXiv} {1707.00062})

\bibitem[\protect\citeauthoryear{{Burgasser} et~al.,}{{Burgasser} et~al.}{2003}]{burg03}
{Burgasser} A.~J.,  et~al., 2003, \mn@doi [\apj] {10.1086/375813}, \href {https://ui.adsabs.harvard.edu/abs/2003ApJ...592.1186B} {592, 1186}

\bibitem[\protect\citeauthoryear{{Burgasser}, {Cruz}  \& {Kirkpatrick}}{{Burgasser} et~al.}{2007}]{burg07}
{Burgasser} A.~J.,  {Cruz} K.~L.,   {Kirkpatrick} J.~D.,  2007, \mn@doi [\apj] {10.1086/510148}, \href {https://ui.adsabs.harvard.edu/abs/2007ApJ...657..494B} {657, 494}

\bibitem[\protect\citeauthoryear{{Burgasser}, {Vrba}, {L{\'e}pine}, {Munn}, {Luginbuhl}, {Henden}, {Guetter}  \& {Canzian}}{{Burgasser} et~al.}{2008}]{burg08}
{Burgasser} A.~J.,  {Vrba} F.~J.,  {L{\'e}pine} S.,  {Munn} J.~A.,  {Luginbuhl} C.~B.,  {Henden} A.~A.,  {Guetter} H.~H.,   {Canzian} B.~C.,  2008, \mn@doi [\apj] {10.1086/523810}, \href {https://ui.adsabs.harvard.edu/abs/2008ApJ...672.1159B} {672, 1159}

\bibitem[\protect\citeauthoryear{{Burrows} et~al.,}{{Burrows} et~al.}{1998}]{burr98}
{Burrows} A.,  et~al., 1998, in {Rebolo} R.,  {Martin} E.~L.,   {Zapatero Osorio} M.~R.,  eds,  Astronomical Society of the Pacific Conference Series Vol. 134, Brown Dwarfs and Extrasolar Planets. p.~354

\bibitem[\protect\citeauthoryear{{Burrows}, {Hubbard}, {Lunine}  \& {Liebert}}{{Burrows} et~al.}{2001}]{burr01}
{Burrows} A.,  {Hubbard} W.~B.,  {Lunine} J.~I.,   {Liebert} J.,  2001, \mn@doi [Reviews of Modern Physics] {10.1103/RevModPhys.73.719}, \href {https://ui.adsabs.harvard.edu/abs/2001RvMP...73..719B} {73, 719}

\bibitem[\protect\citeauthoryear{{Burrows}, {Sudarsky}  \& {Hubeny}}{{Burrows} et~al.}{2006}]{burr06}
{Burrows} A.,  {Sudarsky} D.,   {Hubeny} I.,  2006, \mn@doi [\apj] {10.1086/500293}, \href {https://ui.adsabs.harvard.edu/abs/2006ApJ...640.1063B} {640, 1063}

\bibitem[\protect\citeauthoryear{{Calcaferro}, {Althaus}  \& {C{\'o}rsico}}{{Calcaferro} et~al.}{2018}]{calc18}
{Calcaferro} L.~M.,  {Althaus} L.~G.,   {C{\'o}rsico} A.~H.,  2018, \mn@doi [\aap] {10.1051/0004-6361/201732551}, \href {https://ui.adsabs.harvard.edu/abs/2018A&A...614A..49C} {614, A49}

\bibitem[\protect\citeauthoryear{{Camisassa}, {Althaus}, {C{\'o}rsico}, {Vinyoles}, {Serenelli}, {Isern}, {Miller Bertolami}  \& {Garc{\'\i}a{\textendash}Berro}}{{Camisassa} et~al.}{2016}]{camisassa2016}
{Camisassa} M.~E.,  {Althaus} L.~G.,  {C{\'o}rsico} A.~H.,  {Vinyoles} N.,  {Serenelli} A.~M.,  {Isern} J.,  {Miller Bertolami} M.~M.,   {Garc{\'\i}a{\textendash}Berro} E.,  2016, \mn@doi [\apj] {10.3847/0004-637X/823/2/158}, \href {https://ui.adsabs.harvard.edu/abs/2016ApJ...823..158C} {823, 158}

\bibitem[\protect\citeauthoryear{{Camisassa} et~al.,}{{Camisassa} et~al.}{2019}]{camisassa2019}
{Camisassa} M.~E.,  et~al., 2019, \mn@doi [\aap] {10.1051/0004-6361/201833822}, \href {https://ui.adsabs.harvard.edu/abs/2019A&A...625A..87C} {625, A87}

\bibitem[\protect\citeauthoryear{{Camisassa}, {Baiko}, {Torres}  \& {Rebassa-Mansergas}}{{Camisassa} et~al.}{2024}]{camisassa2024}
{Camisassa} M.,  {Baiko} D.~A.,  {Torres} S.,   {Rebassa-Mansergas} A.,  2024, \mn@doi [\aap] {10.1051/0004-6361/202348344}, \href {https://ui.adsabs.harvard.edu/abs/2024A&A...683A.101C} {683, A101}

\bibitem[\protect\citeauthoryear{{Carollo} et~al.,}{{Carollo} et~al.}{2007}]{caro07}
{Carollo} D.,  et~al., 2007, \mn@doi [\nat] {10.1038/nature06460}, \href {https://ui.adsabs.harvard.edu/abs/2007Natur.450.1020C} {450, 1020}

\bibitem[\protect\citeauthoryear{{Caron}, {Bergeron}, {Blouin}  \& {Leggett}}{{Caron} et~al.}{2023}]{caron2023}
{Caron} A.,  {Bergeron} P.,  {Blouin} S.,   {Leggett} S.~K.,  2023, \mn@doi [\mnras] {10.1093/mnras/stac3733}, \href {https://ui.adsabs.harvard.edu/abs/2023MNRAS.519.4529C} {519, 4529}

\bibitem[\protect\citeauthoryear{{Catal{\'a}n}, {Isern}, {Garc{\'\i}a-Berro}  \& {Ribas}}{{Catal{\'a}n} et~al.}{2008}]{catalan2008}
{Catal{\'a}n} S.,  {Isern} J.,  {Garc{\'\i}a-Berro} E.,   {Ribas} I.,  2008, \mn@doi [\mnras] {10.1111/j.1365-2966.2008.13356.x}, \href {https://ui.adsabs.harvard.edu/abs/2008MNRAS.387.1693C} {387, 1693}

\bibitem[\protect\citeauthoryear{{Chambers} et~al.,}{{Chambers} et~al.}{2016}]{cham16}
{Chambers} K.~C.,  et~al., 2016, \mn@doi [arXiv e-prints] {10.48550/arXiv.1612.05560}, \href {https://ui.adsabs.harvard.edu/abs/2016arXiv161205560C} {p. arXiv:1612.05560}

\bibitem[\protect\citeauthoryear{{Cross} et~al.,}{{Cross} et~al.}{2012}]{cros12}
{Cross} N.~J.~G.,  et~al., 2012, \mn@doi [\aap] {10.1051/0004-6361/201219505}, \href {http://adsabs.harvard.edu/abs/2012A%26A...548A.119C} {548, A119}

\bibitem[\protect\citeauthoryear{{Cukanovaite}, {Tremblay}, {Bergeron}, {Freytag}, {Ludwig}  \& {Steffen}}{{Cukanovaite} et~al.}{2021}]{cukanovaite2021}
{Cukanovaite} E.,  {Tremblay} P.-E.,  {Bergeron} P.,  {Freytag} B.,  {Ludwig} H.-G.,   {Steffen} M.,  2021, \mn@doi [\mnras] {10.1093/mnras/staa3684}, \href {https://ui.adsabs.harvard.edu/abs/2021MNRAS.501.5274C} {501, 5274}

\bibitem[\protect\citeauthoryear{{Day-Jones} et~al.,}{{Day-Jones} et~al.}{2011}]{dayj11}
{Day-Jones} A.~C.,  et~al., 2011, \mn@doi [\mnras] {10.1111/j.1365-2966.2010.17469.x}, \href {https://ui.adsabs.harvard.edu/abs/2011MNRAS.410..705D} {410, 705}

\bibitem[\protect\citeauthoryear{{De Angeli} et~al.,}{{De Angeli} et~al.}{2023}]{BPRP_valid}
{De Angeli} F.,  et~al., 2023, \mn@doi [\aap] {10.1051/0004-6361/202243680}, \href {https://ui.adsabs.harvard.edu/abs/2023A&A...674A...2D} {674, A2}

\bibitem[\protect\citeauthoryear{{Deacon} et~al.,}{{Deacon} et~al.}{2014}]{deac14}
{Deacon} N.~R.,  et~al., 2014, \mn@doi [\apj] {10.1088/0004-637X/792/2/119}, \href {https://ui.adsabs.harvard.edu/abs/2014ApJ...792..119D} {792, 119}

\bibitem[\protect\citeauthoryear{{Dye} et~al.,}{{Dye} et~al.}{2018}]{dye18}
{Dye} S.,  et~al., 2018, \mn@doi [\mnras] {10.1093/mnras/stx2622}, \href {https://ui.adsabs.harvard.edu/abs/2018MNRAS.473.5113D} {473, 5113}

\bibitem[\protect\citeauthoryear{{El-Badry}, {Rix}  \& {Heintz}}{{El-Badry} et~al.}{2021}]{elba21}
{El-Badry} K.,  {Rix} H.-W.,   {Heintz} T.~M.,  2021, \mn@doi [\mnras] {10.1093/mnras/stab323}, \href {https://ui.adsabs.harvard.edu/abs/2021MNRAS.506.2269E} {506, 2269}

\bibitem[\protect\citeauthoryear{{French}, {Casewell}, {Dupuy}, {Debes}, {Manjavacas}, {Martin}  \& {Xu}}{{French} et~al.}{2023}]{Fren23}
{French} J.~R.,  {Casewell} S.~L.,  {Dupuy} T.~J.,  {Debes} J.~H.,  {Manjavacas} E.,  {Martin} E.~C.,   {Xu} S.,  2023, \mn@doi [\mnras] {10.1093/mnras/stac3807}, \href {https://ui.adsabs.harvard.edu/abs/2023MNRAS.519.5008F} {519, 5008}

\bibitem[\protect\citeauthoryear{{GRAVITY Collaboration} et~al.,}{{GRAVITY Collaboration} et~al.}{2019}]{grav19}
{GRAVITY Collaboration} et~al., 2019, \mn@doi [\aap] {10.1051/0004-6361/201935656}, \href {https://ui.adsabs.harvard.edu/abs/2019A&A...625L..10G} {625, L10}

\bibitem[\protect\citeauthoryear{{Gagn{\'e}} et~al.,}{{Gagn{\'e}} et~al.}{2018}]{gagn18}
{Gagn{\'e}} J.,  et~al., 2018, \mn@doi [\apj] {10.3847/1538-4357/aaae09}, \href {https://ui.adsabs.harvard.edu/abs/2018ApJ...856...23G} {856, 23}

\bibitem[\protect\citeauthoryear{{Gaia Collaboration} et~al.,}{{Gaia Collaboration} et~al.}{2016}]{Gaia16}
{Gaia Collaboration} et~al., 2016, \mn@doi [\aap] {10.1051/0004-6361/201629272}, \href {https://ui.adsabs.harvard.edu/abs/2016A&A...595A...1G} {595, A1}

\bibitem[\protect\citeauthoryear{{Gaia Collaboration} et~al.,}{{Gaia Collaboration} et~al.}{2021}]{smar21}
{Gaia Collaboration} et~al., 2021, \mn@doi [\aap] {10.1051/0004-6361/202039498}, \href {https://ui.adsabs.harvard.edu/abs/2021A&A...649A...6G} {649, A6}

\bibitem[\protect\citeauthoryear{{Gaia Collaboration} et~al.,}{{Gaia Collaboration} et~al.}{2023}]{gaia23}
{Gaia Collaboration} et~al., 2023, \mn@doi [\aap] {10.1051/0004-6361/202243940}, \href {https://ui.adsabs.harvard.edu/abs/2023A&A...674A...1G} {674, A1}

\bibitem[\protect\citeauthoryear{{Gentile Fusillo} et~al.,}{{Gentile Fusillo} et~al.}{2019}]{gent19}
{Gentile Fusillo} N.~P.,  et~al., 2019, \mn@doi [\mnras] {10.1093/mnras/sty3016}, \href {https://ui.adsabs.harvard.edu/abs/2019MNRAS.482.4570G} {482, 4570}

\bibitem[\protect\citeauthoryear{{Gentile Fusillo} et~al.,}{{Gentile Fusillo} et~al.}{2021}]{gent21}
{Gentile Fusillo} N.~P.,  et~al., 2021, \mn@doi [\mnras] {10.1093/mnras/stab2672}, \href {https://ui.adsabs.harvard.edu/abs/2021MNRAS.508.3877G} {508, 3877}

\bibitem[\protect\citeauthoryear{{Gizis} \& {Harvin}}{{Gizis} \& {Harvin}}{2006}]{gizi06}
{Gizis} J.~E.,  {Harvin} J.,  2006, \mn@doi [\aj] {10.1086/508514}, \href {http://adsabs.harvard.edu/abs/2006AJ....132.2372G} {132, 2372}

\bibitem[\protect\citeauthoryear{Gálvez, Montes, Fernández-Figueroa, López-Santiago, De~Castro  \& Cornide}{Gálvez et~al.}{2002}]{Galvez_2002}
Gálvez M.~C.,  Montes D.,  Fernández-Figueroa M.~J.,  López-Santiago J.,  De~Castro E.,   Cornide M.,  2002, \mn@doi [Astronomy &amp; Astrophysics] {10.1051/0004-6361:20020644}, 389, 524–536

\bibitem[\protect\citeauthoryear{{Hambly} et~al.,}{{Hambly} et~al.}{2008}]{hamb08}
{Hambly} N.~C.,  et~al., 2008, \mn@doi [\mnras] {10.1111/j.1365-2966.2007.12700.x}, \href {http://adsabs.harvard.edu/abs/2008MNRAS.384..637H} {384, 637}

\bibitem[\protect\citeauthoryear{{Hastings}}{{Hastings}}{1970}]{HASTINGS01041970}
{Hastings} W.~K.,  1970, \mn@doi [Biometrika] {10.1093/biomet/57.1.97}, 57, 97

\bibitem[\protect\citeauthoryear{{Hidalgo} et~al.,}{{Hidalgo} et~al.}{2018}]{hidalgo2018}
{Hidalgo} S.~L.,  et~al., 2018, \mn@doi [\apj] {10.3847/1538-4357/aab158}, \href {https://ui.adsabs.harvard.edu/abs/2018ApJ...856..125H} {856, 125}

\bibitem[\protect\citeauthoryear{{Hook}, {J{\o}rgensen}, {Allington-Smith}, {Davies}, {Metcalfe}, {Murowinski}  \& {Crampton}}{{Hook} et~al.}{2004}]{gmos}
{Hook} I.~M.,  {J{\o}rgensen} I.,  {Allington-Smith} J.~R.,  {Davies} R.~L.,  {Metcalfe} N.,  {Murowinski} R.~G.,   {Crampton} D.,  2004, \mn@doi [\pasp] {10.1086/383624}, \href {https://ui.adsabs.harvard.edu/abs/2004PASP..116..425H} {116, 425}

\bibitem[\protect\citeauthoryear{{Irwin} et~al.,}{{Irwin} et~al.}{2004}]{irwi04}
{Irwin} M.~J.,  et~al., 2004, in {Quinn} P.~J.,  {Bridger} A.,  eds,  \procspie Vol. 5493, Optimizing Scientific Return for Astronomy through Information Technologies. pp 411--422, \mn@doi{10.1117/12.551449}

\bibitem[\protect\citeauthoryear{{Jiang} \& {Tremaine}}{{Jiang} \& {Tremaine}}{2010}]{jian10}
{Jiang} Y.-F.,  {Tremaine} S.,  2010, \mn@doi [\mnras] {10.1111/j.1365-2966.2009.15744.x}, \href {https://ui.adsabs.harvard.edu/abs/2010MNRAS.401..977J} {401, 977}

\bibitem[\protect\citeauthoryear{{Jim{\'e}nez-Esteban}, {Torres}, {Rebassa-Mansergas}, {Skorobogatov}, {Solano}, {Cantero}  \& {Rodrigo}}{{Jim{\'e}nez-Esteban} et~al.}{2018}]{jime18}
{Jim{\'e}nez-Esteban} F.~M.,  {Torres} S.,  {Rebassa-Mansergas} A.,  {Skorobogatov} G.,  {Solano} E.,  {Cantero} C.,   {Rodrigo} C.,  2018, \mn@doi [\mnras] {10.1093/mnras/sty2120}, \href {https://ui.adsabs.harvard.edu/abs/2018MNRAS.480.4505J} {480, 4505}

\bibitem[\protect\citeauthoryear{{Kilic}, {Munn}, {Harris}, {von Hippel}, {Liebert}, {Williams}, {Jeffery}  \& {DeGennaro}}{{Kilic} et~al.}{2017}]{kili17}
{Kilic} M.,  {Munn} J.~A.,  {Harris} H.~C.,  {von Hippel} T.,  {Liebert} J.~W.,  {Williams} K.~A.,  {Jeffery} E.,   {DeGennaro} S.,  2017, \mn@doi [\apj] {10.3847/1538-4357/aa62a5}, \href {https://ui.adsabs.harvard.edu/abs/2017ApJ...837..162K} {837, 162}

\bibitem[\protect\citeauthoryear{{Kilic}, {Bergeron}, {Kosakowski}, {Brown}, {Ag{\"u}eros}  \& {Blouin}}{{Kilic} et~al.}{2020}]{kili20}
{Kilic} M.,  {Bergeron} P.,  {Kosakowski} A.,  {Brown} W.~R.,  {Ag{\"u}eros} M.~A.,   {Blouin} S.,  2020, \mn@doi [\apj] {10.3847/1538-4357/ab9b8d}, \href {https://ui.adsabs.harvard.edu/abs/2020ApJ...898...84K} {898, 84}

\bibitem[\protect\citeauthoryear{{Kirkpatrick}, {Henry}  \& {Liebert}}{{Kirkpatrick} et~al.}{1993}]{kirk93}
{Kirkpatrick} J.~D.,  {Henry} T.~J.,   {Liebert} J.,  1993, \mn@doi [\apj] {10.1086/172480}, \href {https://ui.adsabs.harvard.edu/abs/1993ApJ...406..701K} {406, 701}

\bibitem[\protect\citeauthoryear{{Kirkpatrick} et~al.,}{{Kirkpatrick} et~al.}{1999a}]{kirk99}
{Kirkpatrick} J.~D.,  et~al., 1999a, \mn@doi [\apj] {10.1086/307414}, \href {https://ui.adsabs.harvard.edu/abs/1999ApJ...519..802K} {519, 802}

\bibitem[\protect\citeauthoryear{{Kirkpatrick}, {Allard}, {Bida}, {Zuckerman}, {Becklin}, {Chabrier}  \& {Baraffe}}{{Kirkpatrick} et~al.}{1999b}]{kirk99b}
{Kirkpatrick} J.~D.,  {Allard} F.,  {Bida} T.,  {Zuckerman} B.,  {Becklin} E.~E.,  {Chabrier} G.,   {Baraffe} I.,  1999b, \mn@doi [\apj] {10.1086/307380}, \href {https://ui.adsabs.harvard.edu/abs/1999ApJ...519..834K} {519, 834}

\bibitem[\protect\citeauthoryear{{Kirkpatrick} et~al.,}{{Kirkpatrick} et~al.}{2010}]{kirk10}
{Kirkpatrick} J.~D.,  et~al., 2010, \mn@doi [\apjs] {10.1088/0067-0049/190/1/100}, \href {https://ui.adsabs.harvard.edu/abs/2010ApJS..190..100K} {190, 100}

\bibitem[\protect\citeauthoryear{{Kirkpatrick} et~al.,}{{Kirkpatrick} et~al.}{2014}]{kirk14}
{Kirkpatrick} J.~D.,  et~al., 2014, \mn@doi [\apj] {10.1088/0004-637X/783/2/122}, \href {https://ui.adsabs.harvard.edu/abs/2014ApJ...783..122K} {783, 122}

\bibitem[\protect\citeauthoryear{{Kirkpatrick} et~al.,}{{Kirkpatrick} et~al.}{2021}]{kirk21}
{Kirkpatrick} J.~D.,  et~al., 2021, \mn@doi [\apjs] {10.3847/1538-4365/abd107}, \href {https://ui.adsabs.harvard.edu/abs/2021ApJS..253....7K} {253, 7}

\bibitem[\protect\citeauthoryear{{Kowalski} \& {Saumon}}{{Kowalski} \& {Saumon}}{2006}]{saumon2006}
{Kowalski} P.~M.,  {Saumon} D.,  2006, \mn@doi [\apjl] {10.1086/509723}, \href {https://ui.adsabs.harvard.edu/abs/2006ApJ...651L.137K} {651, L137}

\bibitem[\protect\citeauthoryear{{Labrie}, {Cardenes}, {Anderson}, {Simpson}  \& {Turner}}{{Labrie} et~al.}{2020}]{dragons}
{Labrie} K.,  {Cardenes} R.,  {Anderson} K.,  {Simpson} C.,   {Turner} J.,  2020, in {Ballester} P.,  {Ibsen} J.,  {Solar} M.,   {Shortridge} K.,  eds,  Astronomical Society of the Pacific Conference Series Vol. 522, Astronomical Data Analysis Software and Systems XXVII. p.~583

\bibitem[\protect\citeauthoryear{{Lallement}, {Babusiaux}, {Vergely}, {Katz}, {Arenou}, {Valette}, {Hottier}  \& {Capitanio}}{{Lallement} et~al.}{2019}]{lallement2019}
{Lallement} R.,  {Babusiaux} C.,  {Vergely} J.~L.,  {Katz} D.,  {Arenou} F.,  {Valette} B.,  {Hottier} C.,   {Capitanio} L.,  2019, \mn@doi [\aap] {10.1051/0004-6361/201834695}, \href {https://ui.adsabs.harvard.edu/abs/2019A&A...625A.135L} {625, A135}

\bibitem[\protect\citeauthoryear{{Linsky}}{{Linsky}}{1969}]{lins69}
{Linsky} J.~L.,  1969, \mn@doi [\apj] {10.1086/150030}, \href {http://adsabs.harvard.edu/abs/1969ApJ...156..989L} {156, 989}

\bibitem[\protect\citeauthoryear{{Luhman}, {Burgasser}  \& {Bochanski}}{{Luhman} et~al.}{2011}]{luhm11}
{Luhman} K.~L.,  {Burgasser} A.~J.,   {Bochanski} J.~J.,  2011, \mn@doi [\apjl] {10.1088/2041-8205/730/1/L9}, \href {https://ui.adsabs.harvard.edu/abs/2011ApJ...730L...9L} {730, L9}

\bibitem[\protect\citeauthoryear{{Marley} et~al.,}{{Marley} et~al.}{2021}]{marl21}
{Marley} M.~S.,  et~al., 2021, \mn@doi [\apj] {10.3847/1538-4357/ac141d}, \href {https://ui.adsabs.harvard.edu/abs/2021ApJ...920...85M} {920, 85}

\bibitem[\protect\citeauthoryear{{Marocco} et~al.,}{{Marocco} et~al.}{2021}]{maro21}
{Marocco} F.,  et~al., 2021, \mn@doi [\apjs] {10.3847/1538-4365/abd805}, \href {https://ui.adsabs.harvard.edu/abs/2021ApJS..253....8M} {253, 8}

\bibitem[\protect\citeauthoryear{{Mart{\'\i}n}, {Delfosse}, {Basri}, {Goldman}, {Forveille}  \& {Zapatero Osorio}}{{Mart{\'\i}n} et~al.}{1999}]{mart99}
{Mart{\'\i}n} E.~L.,  {Delfosse} X.,  {Basri} G.,  {Goldman} B.,  {Forveille} T.,   {Zapatero Osorio} M.~R.,  1999, \mn@doi [\aj] {10.1086/301107}, \href {https://ui.adsabs.harvard.edu/abs/1999AJ....118.2466M} {118, 2466}

\bibitem[\protect\citeauthoryear{{Meisner} et~al.,}{{Meisner} et~al.}{2020}]{meis20}
{Meisner} A.~M.,  et~al., 2020, \mn@doi [\apj] {10.3847/1538-4357/aba633}, \href {https://ui.adsabs.harvard.edu/abs/2020ApJ...899..123M} {899, 123}

\bibitem[\protect\citeauthoryear{{Meisner} et~al.,}{{Meisner} et~al.}{2021}]{meis21}
{Meisner} A.~M.,  et~al., 2021, \mn@doi [\apj] {10.3847/1538-4357/ac013c}, \href {https://ui.adsabs.harvard.edu/abs/2021ApJ...915..120M} {915, 120}

\bibitem[\protect\citeauthoryear{{Metropolis}, {Rosenbluth}, {Rosenbluth}, {Teller}  \& {Teller}}{{Metropolis} et~al.}{1953}]{1953JChPh..21.1087M}
{Metropolis} N.,  {Rosenbluth} A.~W.,  {Rosenbluth} M.~N.,  {Teller} A.~H.,   {Teller} E.,  1953, \mn@doi [\jcp] {10.1063/1.1699114}, \href {http://adsabs.harvard.edu/abs/1953JChPh..21.1087M} {21, 1087}

\bibitem[\protect\citeauthoryear{{Minniti} et~al.,}{{Minniti} et~al.}{2010}]{minn10}
{Minniti} D.,  et~al., 2010, \mn@doi [\na] {10.1016/j.newast.2009.12.002}, \href {https://ui.adsabs.harvard.edu/abs/2010NewA...15..433M} {15, 433}

\bibitem[\protect\citeauthoryear{{Montegriffo} et~al.,}{{Montegriffo} et~al.}{2023}]{BPRP_calib}
{Montegriffo} P.,  et~al., 2023, \mn@doi [\aap] {10.1051/0004-6361/202243880}, \href {https://ui.adsabs.harvard.edu/abs/2023A&A...674A...3M} {674, A3}

\bibitem[\protect\citeauthoryear{{Mukherjee} et~al.,}{{Mukherjee} et~al.}{2024}]{mukh24}
{Mukherjee} S.,  et~al., 2024, \mn@doi [\apj] {10.3847/1538-4357/ad18c2}, \href {https://ui.adsabs.harvard.edu/abs/2024ApJ...963...73M} {963, 73}

\bibitem[\protect\citeauthoryear{{Nakajima}, {Oppenheimer}, {Kulkarni}, {Golimowski}, {Matthews}  \& {Durrance}}{{Nakajima} et~al.}{1995}]{naka95}
{Nakajima} T.,  {Oppenheimer} B.~R.,  {Kulkarni} S.~R.,  {Golimowski} D.~A.,  {Matthews} K.,   {Durrance} S.~T.,  1995, \mn@doi [\nat] {10.1038/378463a0}, \href {https://ui.adsabs.harvard.edu/abs/1995Natur.378..463N} {378, 463}

\bibitem[\protect\citeauthoryear{{O'Brien} et~al.,}{{O'Brien} et~al.}{2024}]{obrien2024}
{O'Brien} M.~W.,  et~al., 2024, \mn@doi [\mnras] {10.1093/mnras/stad3773}, \href {https://ui.adsabs.harvard.edu/abs/2024MNRAS.527.8687O} {527, 8687}

\bibitem[\protect\citeauthoryear{{Pietrinferni} et~al.,}{{Pietrinferni} et~al.}{2021}]{pietrinferni2021}
{Pietrinferni} A.,  et~al., 2021, \mn@doi [\apj] {10.3847/1538-4357/abd4d5}, \href {https://ui.adsabs.harvard.edu/abs/2021ApJ...908..102P} {908, 102}

\bibitem[\protect\citeauthoryear{{Raddi} et~al.,}{{Raddi} et~al.}{2022}]{raddi2022}
{Raddi} R.,  et~al., 2022, \mn@doi [\aap] {10.1051/0004-6361/202141837}, \href {https://ui.adsabs.harvard.edu/abs/2022A&A...658A..22R} {658, A22}

\bibitem[\protect\citeauthoryear{{Reddy}, {Lambert}  \& {Allende Prieto}}{{Reddy} et~al.}{2006}]{redd06}
{Reddy} B.~E.,  {Lambert} D.~L.,   {Allende Prieto} C.,  2006, \mn@doi [\mnras] {10.1111/j.1365-2966.2006.10148.x}, \href {https://ui.adsabs.harvard.edu/abs/2006MNRAS.367.1329R} {367, 1329}

\bibitem[\protect\citeauthoryear{{Reyl{\'e}}}{{Reyl{\'e}}}{2018}]{reyl18}
{Reyl{\'e}} C.,  2018, \mn@doi [\aap] {10.1051/0004-6361/201834082}, \href {https://ui.adsabs.harvard.edu/abs/2018A&A...619L...8R} {619, L8}

\bibitem[\protect\citeauthoryear{{Romero}, {Campos}  \& {Kepler}}{{Romero} et~al.}{2015}]{romero2015}
{Romero} A.~D.,  {Campos} F.,   {Kepler} S.~O.,  2015, \mn@doi [\mnras] {10.1093/mnras/stv848}, \href {https://ui.adsabs.harvard.edu/abs/2015MNRAS.450.3708R} {450, 3708}

\bibitem[\protect\citeauthoryear{{Ruiz}, {Leggett}  \& {Allard}}{{Ruiz} et~al.}{1997}]{ruiz97}
{Ruiz} M.~T.,  {Leggett} S.~K.,   {Allard} F.,  1997, \mn@doi [\apjl] {10.1086/311070}, \href {https://ui.adsabs.harvard.edu/abs/1997ApJ...491L.107R} {491, L107}

\bibitem[\protect\citeauthoryear{{Salaris} \& {Cassisi}}{{Salaris} \& {Cassisi}}{2005}]{sala05}
{Salaris} M.,  {Cassisi} S.,  2005, {Evolution of Stars and Stellar Populations}

\bibitem[\protect\citeauthoryear{{Saydjari} et~al.,}{{Saydjari} et~al.}{2023}]{decaps2}
{Saydjari} A.~K.,  et~al., 2023, \mn@doi [\apjs] {10.3847/1538-4365/aca594}, \href {https://ui.adsabs.harvard.edu/abs/2023ApJS..264...28S} {264, 28}

\bibitem[\protect\citeauthoryear{{Schlafly} et~al.,}{{Schlafly} et~al.}{2018}]{decaps1}
{Schlafly} E.~F.,  et~al., 2018, \mn@doi [\apjs] {10.3847/1538-4365/aaa3e2}, \href {https://ui.adsabs.harvard.edu/abs/2018ApJS..234...39S} {234, 39}

\bibitem[\protect\citeauthoryear{{Serenelli}, {Althaus}, {Rohrmann}  \& {Benvenuto}}{{Serenelli} et~al.}{2002}]{serenelli2002}
{Serenelli} A.~M.,  {Althaus} L.~G.,  {Rohrmann} R.~D.,   {Benvenuto} O.~G.,  2002, \mn@doi [\mnras] {10.1046/j.1365-8711.2002.05994.x}, \href {https://ui.adsabs.harvard.edu/abs/2002MNRAS.337.1091S} {337, 1091}

\bibitem[\protect\citeauthoryear{{Skrutskie} et~al.,}{{Skrutskie} et~al.}{2006}]{skru06}
{Skrutskie} M.~F.,  et~al., 2006, \mn@doi [\aj] {10.1086/498708}, \href {https://ui.adsabs.harvard.edu/abs/2006AJ....131.1163S} {131, 1163}

\bibitem[\protect\citeauthoryear{{Smith} et~al.,}{{Smith} et~al.}{2018}]{smit18}
{Smith} L.~C.,  et~al., 2018, \mn@doi [\mnras] {10.1093/mnras/stx2789}, \href {https://ui.adsabs.harvard.edu/abs/2018MNRAS.474.1826S} {474, 1826}

\bibitem[\protect\citeauthoryear{{Steele}, {Burleigh}, {Farihi}, {G{\"a}nsicke}, {Jameson}, {Dobbie}  \& {Barstow}}{{Steele} et~al.}{2009}]{stee09}
{Steele} P.~R.,  {Burleigh} M.~R.,  {Farihi} J.,  {G{\"a}nsicke} B.~T.,  {Jameson} R.~F.,  {Dobbie} P.~D.,   {Barstow} M.~A.,  2009, \mn@doi [\aap] {10.1051/0004-6361/200911694}, \href {https://ui.adsabs.harvard.edu/abs/2009A&A...500.1207S} {500, 1207}

\bibitem[\protect\citeauthoryear{{Tody}}{{Tody}}{1993}]{tody1993}
{Tody} D.,  1993, Astronomical Data Analysis Software and Systems II, A.S.P. Conference Series, \href {https://articles.adsabs.harvard.edu/pdf/1993ASPC...52..173T} {52}

\bibitem[\protect\citeauthoryear{{Tonry} \& {Davis}}{{Tonry} \& {Davis}}{1979}]{tonryanddavis1979}
{Tonry} J.,  {Davis} M.,  1979, \mn@doi [\aj] {10.1086/112569}, \href {https://articles.adsabs.harvard.edu/pdf/1979AJ.....84.1511T} {84, 1511}

\bibitem[\protect\citeauthoryear{{Tremblay}, {Bergeron}  \& {Gianninas}}{{Tremblay} et~al.}{2011}]{tremblay2011}
{Tremblay} P.~E.,  {Bergeron} P.,   {Gianninas} A.,  2011, \mn@doi [\apj] {10.1088/0004-637X/730/2/128}, \href {https://ui.adsabs.harvard.edu/abs/2011ApJ...730..128T} {730, 128}

\bibitem[\protect\citeauthoryear{{Tremblay}, {Ludwig}, {Steffen}  \& {Freytag}}{{Tremblay} et~al.}{2013}]{tremblay2013}
{Tremblay} P.~E.,  {Ludwig} H.~G.,  {Steffen} M.,   {Freytag} B.,  2013, \mn@doi [\aap] {10.1051/0004-6361/201322318}, \href {https://ui.adsabs.harvard.edu/abs/2013A&A...559A.104T} {559, A104}

\bibitem[\protect\citeauthoryear{{Tremblay}, {Cummings}, {Kalirai}, {G{\"a}nsicke}, {Gentile-Fusillo}  \& {Raddi}}{{Tremblay} et~al.}{2016}]{tremblay2016}
{Tremblay} P.~E.,  {Cummings} J.,  {Kalirai} J.~S.,  {G{\"a}nsicke} B.~T.,  {Gentile-Fusillo} N.,   {Raddi} R.,  2016, \mn@doi [\mnras] {10.1093/mnras/stw1447}, \href {https://ui.adsabs.harvard.edu/abs/2016MNRAS.461.2100T} {461, 2100}

\bibitem[\protect\citeauthoryear{{Valenti} et~al.,}{{Valenti} et~al.}{2016}]{vale16}
{Valenti} E.,  et~al., 2016, \mn@doi [\aap] {10.1051/0004-6361/201527500}, \href {https://ui.adsabs.harvard.edu/abs/2016A&A...587L...6V} {587, L6}

\bibitem[\protect\citeauthoryear{{Vernet} et~al.,}{{Vernet} et~al.}{2011}]{vern11}
{Vernet} J.,  et~al., 2011, \mn@doi [\aap] {10.1051/0004-6361/201117752}, \href {https://ui.adsabs.harvard.edu/abs/2011A&A...536A.105V} {536, A105}

\bibitem[\protect\citeauthoryear{{Weinberg}, {Shapiro}  \& {Wasserman}}{{Weinberg} et~al.}{1987}]{wein87}
{Weinberg} M.~D.,  {Shapiro} S.~L.,   {Wasserman} I.,  1987, \mn@doi [\apj] {10.1086/164883}, \href {https://ui.adsabs.harvard.edu/abs/1987ApJ...312..367W} {312, 367}

\bibitem[\protect\citeauthoryear{{West}, {Hawley}, {Bochanski}, {Covey}, {Reid}, {Dhital}, {Hilton}  \& {Masuda}}{{West} et~al.}{2008}]{west08}
{West} A.~A.,  {Hawley} S.~L.,  {Bochanski} J.~J.,  {Covey} K.~R.,  {Reid} I.~N.,  {Dhital} S.,  {Hilton} E.~J.,   {Masuda} M.,  2008, \mn@doi [\aj] {10.1088/0004-6256/135/3/785}, \href {https://ui.adsabs.harvard.edu/abs/2008AJ....135..785W} {135, 785}

\bibitem[\protect\citeauthoryear{{Witte}, {Helling}, {Barman}, {Heidrich}  \& {Hauschildt}}{{Witte} et~al.}{2011}]{witt11}
{Witte} S.,  {Helling} C.,  {Barman} T.,  {Heidrich} N.,   {Hauschildt} P.~H.,  2011, \mn@doi [\aap] {10.1051/0004-6361/201014105}, \href {https://ui.adsabs.harvard.edu/abs/2011A&A...529A..44W} {529, A44}

\bibitem[\protect\citeauthoryear{{Wright} et~al.,}{{Wright} et~al.}{2010}]{wrig10}
{Wright} E.~L.,  et~al., 2010, \mn@doi [\aj] {10.1088/0004-6256/140/6/1868}, \href {https://ui.adsabs.harvard.edu/abs/2010AJ....140.1868W} {140, 1868}

\bibitem[\protect\citeauthoryear{{York} et~al.,}{{York} et~al.}{2000}]{york00}
{York} D.~G.,  et~al., 2000, \mn@doi [\aj] {10.1086/301513}, \href {https://ui.adsabs.harvard.edu/abs/2000AJ....120.1579Y} {120, 1579}

\bibitem[\protect\citeauthoryear{{Zhang}}{{Zhang}}{2019}]{prime7}
{Zhang} Z.,  2019, \mn@doi [\mnras] {10.1093/mnras/stz2196}, \href {https://ui.adsabs.harvard.edu/abs/2019MNRAS.489.1423Z} {489, 1423}

\bibitem[\protect\citeauthoryear{{Zhang} et~al.,}{{Zhang} et~al.}{2013}]{zhan13}
{Zhang} Z.~H.,  et~al., 2013, \mn@doi [\mnras] {10.1093/mnras/stt1030}, \href {https://ui.adsabs.harvard.edu/abs/2013MNRAS.434.1005Z} {434, 1005}

\bibitem[\protect\citeauthoryear{{Zhang} et~al.,}{{Zhang} et~al.}{2017a}]{prime1}
{Zhang} Z.~H.,  et~al., 2017a, \mn@doi [\mnras] {10.1093/mnras/stw2438}, \href {https://ui.adsabs.harvard.edu/abs/2017MNRAS.464.3040Z} {464, 3040}

\bibitem[\protect\citeauthoryear{{Zhang}, {Homeier}, {Pinfield}, {Lodieu}, {Jones}, {Allard}  \& {Pavlenko}}{{Zhang} et~al.}{2017b}]{prime2}
{Zhang} Z.~H.,  {Homeier} D.,  {Pinfield} D.~J.,  {Lodieu} N.,  {Jones} H.~R.~A.,  {Allard} F.,   {Pavlenko} Y.~V.,  2017b, \mn@doi [\mnras] {10.1093/mnras/stx350}, \href {https://ui.adsabs.harvard.edu/abs/2017MNRAS.468..261Z} {468, 261}

\bibitem[\protect\citeauthoryear{{Zhang} et~al.,}{{Zhang} et~al.}{2018a}]{prime3}
{Zhang} Z.~H.,  et~al., 2018a, \mn@doi [\mnras] {10.1093/mnras/sty1352}, \href {https://ui.adsabs.harvard.edu/abs/2018MNRAS.479.1383Z} {479, 1383}

\bibitem[\protect\citeauthoryear{{Zhang} et~al.,}{{Zhang} et~al.}{2018b}]{prime4}
{Zhang} Z.~H.,  et~al., 2018b, \mn@doi [\mnras] {10.1093/mnras/sty2054}, \href {https://ui.adsabs.harvard.edu/abs/2018MNRAS.480.5447Z} {480, 5447}

\bibitem[\protect\citeauthoryear{{Zhang}, {Burgasser}, {G{\'a}lvez-Ortiz}, {Lodieu}, {Zapatero Osorio}, {Pinfield}  \& {Allard}}{{Zhang} et~al.}{2019a}]{prime6}
{Zhang} Z.~H.,  {Burgasser} A.~J.,  {G{\'a}lvez-Ortiz} M.~C.,  {Lodieu} N.,  {Zapatero Osorio} M.~R.,  {Pinfield} D.~J.,   {Allard} F.,  2019a, \mn@doi [\mnras] {10.1093/mnras/stz777}, \href {https://ui.adsabs.harvard.edu/abs/2019MNRAS.486.1260Z} {486, 1260}

\bibitem[\protect\citeauthoryear{{Zhang}, {Burgasser}  \& {Smith}}{{Zhang} et~al.}{2019b}]{prime5}
{Zhang} Z.~H.,  {Burgasser} A.~J.,   {Smith} L.~C.,  2019b, \mn@doi [\mnras] {10.1093/mnras/stz659}, \href {https://ui.adsabs.harvard.edu/abs/2019MNRAS.486.1840Z} {486, 1840}

\bibitem[\protect\citeauthoryear{{Zhang} et~al.,}{{Zhang} et~al.}{2020}]{zhan20}
{Zhang} Z.,  et~al., 2020, \mn@doi [\apj] {10.3847/1538-4357/ab765c}, \href {https://ui.adsabs.harvard.edu/abs/2020ApJ...891..171Z} {891, 171}

\bibitem[\protect\citeauthoryear{{Zubiaur}, {Raddi}  \& {Torres}}{{Zubiaur} et~al.}{2024}]{zubi24}
{Zubiaur} A.,  {Raddi} R.,   {Torres} S.,  2024, \mn@doi [\aap] {10.1051/0004-6361/202449223}, \href {https://ui.adsabs.harvard.edu/abs/2024A&A...687A.286Z} {687, A286}

\makeatother
\end{thebibliography}




\appendix
\section{VVV~1256$-$62AB's orbit compared to the  Milky Way}
\begin{figure*}
    \centering
    \includegraphics[width=\textwidth]{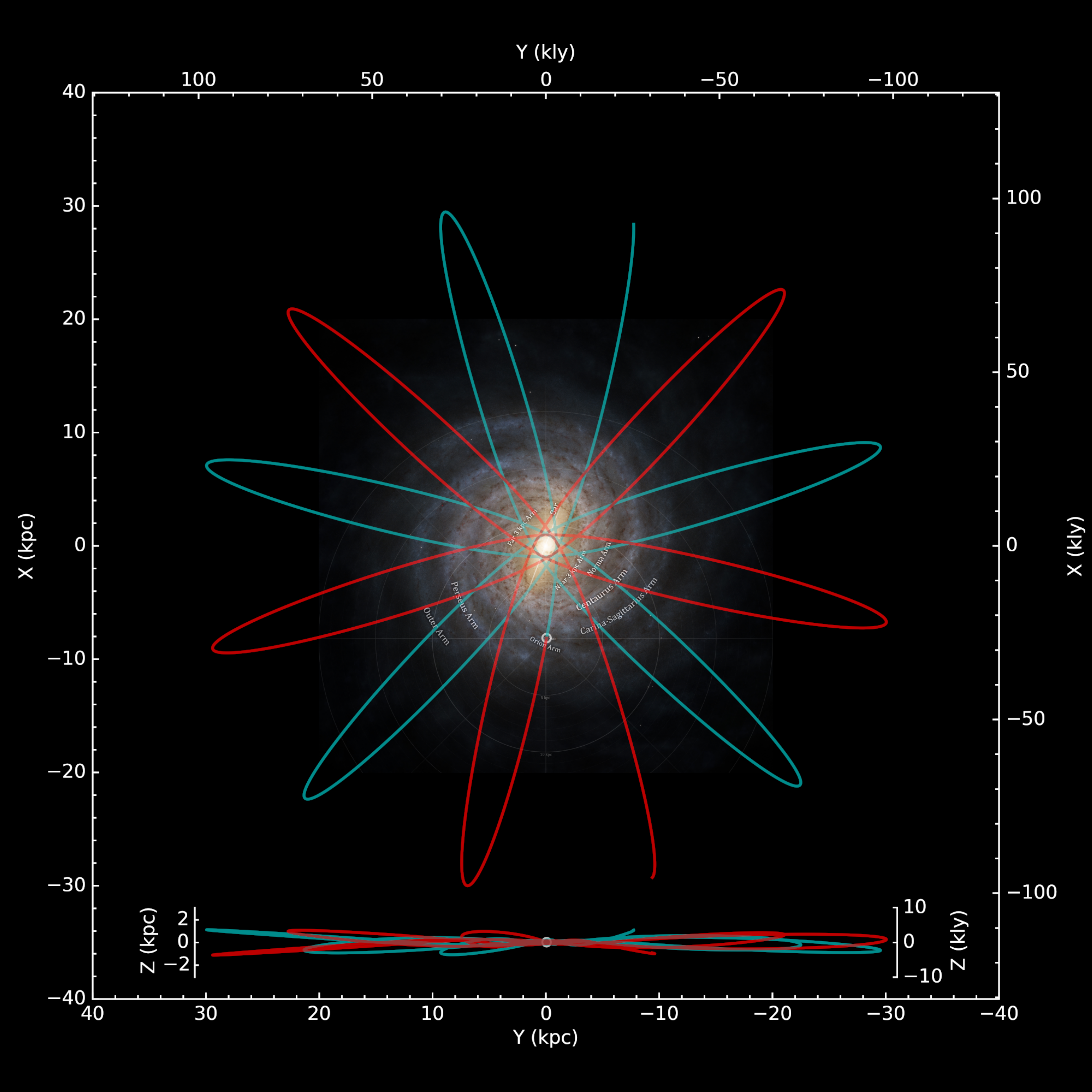} 
\caption{VVV~1256$-$62AB's average orbit from the past 2Gyr (cyan curves) to the future 2Gyr (red curves) compared to the spiral structure of the Milky Way (based on {\sl Gaia} DR3; Credits: ESA/Gaia/DPAC, Stefan Payne-Wardenaar, CC BY-SA 4.0 IGO). Its current location is indicated with a white circle near the Sun. The edge-on orbit view in [Y, Z] space is plotted on the same scale and shown at the bottom. }
    \label{forbitmw}
\end{figure*}

To better visualize the Galactic orbit of VVV~1256$-$62AB we compared its average orbit to the face-on spiral structure of the Milky Way based on {\sl Gaia} DR3 in Fig. \ref{forbitmw}. The edge-on orbit view in [Y, Z] space shows VVV~1256$-$62AB have very flat orbit with a diameter to thickness ratio of about 28.  

\bsp	
\label{lastpage}
\end{document}